\pdfoutput=1
\documentclass[11pt,twoside,a4paper,cmspaper,final,collab]{cms-tdr}
\def\svnVersion{7be3341}\def\svnDate{2025/12/02}\def\cmsCernNoTag{CERN-EP-2025-007}\def\cmsCernDate{\today}\def\cmsMessage{Published in Physical Review D as \href{http://dx.doi.org/10.1103/6s8n-pvmy}{\doi{10.1103/6s8n-pvmy}.}}
\begin{document}\cmsNoteHeader{HIG-23-011}

\newlength\cmsFigWidth
\ifthenelse{\boolean{cms@external}}{\setlength\cmsFigWidth{0.35\textwidth}}{\setlength\cmsFigWidth{0.5\textwidth}}
\newlength\cmsFigWidthF
\ifthenelse{\boolean{cms@external}}{\setlength\cmsFigWidthF{0.5\textwidth}}{\setlength\cmsFigWidthF{0.6\textwidth}}
\ifthenelse{\boolean{cms@external}}{\providecommand{\cmsLeft}{upper\xspace}}{\providecommand{\cmsLeft}{left\xspace}}
\ifthenelse{\boolean{cms@external}}{\providecommand{\cmsRight}{lower\xspace}}{\providecommand{\cmsRight}{right\xspace}}
\newlength\cmsTabSkip\setlength\cmsTabSkip{1.2ex}
\providecommand{\cmsTable}[1]{\resizebox{\textwidth}{!}{#1}}
\newcommand{\jhugen}{\textsc{JHUGen}\xspace}
\newcommand{\mela}{\textsc{mela}\xspace}
\newcommand{\ZZ}{\ensuremath{\PZ\PZ}\xspace}
\newcommand{\HVV}{\ensuremath{\PH\PV\PV}\xspace}
\newcommand{\HtoVV}{\ensuremath{\PH\to\PV\PV}\xspace}
\newcommand{\Hgg}{\ensuremath{\PH\Pg\Pg}\xspace}
\newcommand{\VBF}{\ensuremath{\mathrm{VBF}}\xspace}
\newcommand{\ZH}{\ensuremath{\PZ\PH}\xspace}
\newcommand{\ggZH}{\ensuremath{\Pg\Pg\to\PZ\PH}\xspace}
\newcommand{\xsecGammaH}{\ensuremath{\sigma_{\PGg\PH}}\xspace}
\newcommand{\ttH}{\ensuremath{\PQt\PAQt\PH}\xspace}
\newcommand{\bbH}{\ensuremath{\PQb\PAQb\PH}\xspace}
\newcommand{\ggH}{\ensuremath{\Pg\Pg\PH}\xspace}
\newcommand{\ZX}{\ensuremath{\PZ+\PX}\xspace}
\newcommand{\Hboson}{\ensuremath{\PH} boson\xspace}
\newcommand{\Hell}{\ensuremath{\PH\to4\Pell}\xspace}
\newcommand{\GH}{\ensuremath{\Gamma_\PH}\xspace}
\newcommand{\mH}{\ensuremath{m_{\PH}}\xspace}
\newcommand{\mell}{\ensuremath{m_{4\Pell}}\xspace}
\newcommand{\Psig}{\ensuremath{\mathcal{P}_\text{sig}}\xspace}
\newcommand{\Pbkg}{\ensuremath{\mathcal{P}_\text{bkg}}\xspace}
\newcommand{\Dbkg}{\ensuremath{{\mathcal{D}}_{\text{bkg}}}\xspace}
\newcommand{\gammaH}{\ensuremath{\PGg \PH}\xspace}
\newcommand{\HZy}{\ensuremath{\PH\PZ\PGg}\xspace}
\newcommand{\Hyy}{\ensuremath{\PH\PGg\PGg}\xspace}
\newcommand{\fxfx}{{\textsc{FxFx}}\xspace}
\newcommand{\mpnet}{\ensuremath{M_{\textrm{PNet}}}\xspace}
\newcommand{\Xbb}{\ensuremath{\PX\to\bbbar}\xspace}
\newcommand{\Zjets}{\ensuremath{\PZ+\text{jets}}\xspace}
\newcommand{\Wjets}{\ensuremath{\PW+\text{jets}}\xspace}
\newcommand{\Vgamma}{\ensuremath{\PV+\PGg}\xspace}
\newcommand{\rpf}{\ensuremath{R_\text{{P/F}}}\xspace}
\newcommand{\Zgamma}{\ensuremath{\PZ+\PGg}\xspace}
\newcommand{\Wgamma}{\ensuremath{\PW+\PGg}\xspace}
\newcommand{\czy}{\ensuremath{c_{z\PGg}}\xspace}
\newcommand{\cyy}{\ensuremath{c_{\PGg\PGg}}\xspace}
\newcommand{\tczy}{\ensuremath{\widetilde{c}_{z\gamma}}\xspace}
\newcommand{\tcyy}{\ensuremath{\widetilde{c}_{\gamma\gamma}}\xspace}
\newcommand{\qqbarH}{\ensuremath{\PQq\PAQq\PH}\xspace}
\newcommand{\Htobbbar}{\ensuremath{\PH\to\PQb\PAQb}\xspace}
\newcommand{\Htoqqbar}{\ensuremath{\PH\to\PQq\PAQq}\xspace}
\newcommand{\CqW}{\ensuremath{C^{\PQq\PW}}\xspace}
\newcommand{\CqB}{\ensuremath{C^{\PQq\PB}}\xspace}
\newcommand{\Hqq}{\ensuremath{\PH\PQq\PQq}\xspace}
\newcommand{\HfourL}{\ensuremath{\PH\to 4\Pell}\xspace}
\newcommand{\PQCD}{\ensuremath{P(\text{QCD})}\xspace}
\newcommand{\PXbb}{\ensuremath{P(\Xbb)}\xspace}
\newcommand{\kappau}{\ensuremath{\kappa_{\PQu}}\xspace}
\newcommand{\kappad}{\ensuremath{\kappa_{\PQd}}\xspace}
\newcommand{\kappas}{\ensuremath{\kappa_{\PQs}}\xspace}
\newcommand{\kappac}{\ensuremath{\kappa_{\PQc}}\xspace}
\newcommand{\kappab}{\ensuremath{\kappa_{\PQb}}\xspace}
\newcommand{\kappat}{\ensuremath{\kappa_{\PQt}}\xspace}
\newcommand{\kappaq}{\ensuremath{\kappa_{\PQq}}\xspace}
\newcommand{\kappaall}{\ensuremath{\kappa_{\PQu,\PQd,\PQs,\PQc,\PQb}}\xspace}
\newcommand{\Rgg}{\ensuremath{R_{\Pg\Pg}}\xspace}
\newcommand{\GHBSM}{\ensuremath{\Gamma^\text{BSM}_\PH}\xspace}
\newcommand{\kappavvprime}{\ensuremath{\kappa_{\PV\PV^\prime}}\xspace}
\newcommand{\HtoVVprime}{\ensuremath{\HtoVV^\prime}\xspace}
\newcommand{\Htogg}{\ensuremath{\PH\to\Pg\Pg}\xspace}
\newcommand{\kappaZZ}{\ensuremath{\kappa_{\ZZ}}\xspace}
\newcommand{\kappaWW}{\ensuremath{\kappa_{\PW\PW}}\xspace}
\newcommand{\barkappau}{\ensuremath{\overline{\kappa}_{\PQu}}\xspace}
\newcommand{\barkappad}{\ensuremath{\overline{\kappa}_{\PQd}}\xspace}
\newcommand{\barkappas}{\ensuremath{\overline{\kappa}_{\PQs}}\xspace}
\newcommand{\barkappac}{\ensuremath{\overline{\kappa}_{\PQc}}\xspace}
\newcommand{\barkappab}{\ensuremath{\overline{\kappa}_{\PQb}}\xspace}
\newcommand{\barkappat}{\ensuremath{\overline{\kappa}_{\PQt}}\xspace}
\newcommand{\barkappaq}{\ensuremath{\overline{\kappa}_{\PQq}}\xspace}
\newcommand{\Vjets}{\ensuremath{\PV+\text{jets}}\xspace}
\newcommand{\CP}{\ensuremath{C\mkern-3mu P}\xspace}

\cmsNoteHeader{HIG-23-011}
\title{Search for \texorpdfstring{\gammaH}{gamma H} production and constraints on the Yukawa couplings of light quarks to the Higgs boson}

\date{\today}

\abstract{
A search for \gammaH production is performed with data from the CMS experiment at the LHC  corresponding to an integrated luminosity of 138\fbinv at a proton-proton center-of-mass collision energy of 13\TeV. The analysis focuses on the topology of a boosted Higgs boson recoiling against a high-energy photon.  The final states of $\PH\to\bbbar$ and \Hell are analyzed.  This study examines effective \HZy and \Hyy anomalous couplings within the context of an effective field theory. In this approach, the production cross section is constrained to be $\sigma_{\gammaH}<16.4\unit{fb}$ at 95\% confidence level (\CL). Simultaneous constraints on four anomalous couplings involving \HZy and \Hyy are provided. Additionally, the production rate for \Hell is examined to assess potential enhancements in the Yukawa couplings  between light quarks and the Higgs boson.  Assuming the standard model values for the Yukawa couplings of the bottom and top quarks, the following simultaneous constraints are obtained:   $\kappa_\PQu=(0.0\pm1.5)\times10^{3}$, $\kappa_\PQd=(0.0^{+6.7}_{-6.8})\times10^{2}$, $\kappa_\PQs=0^{+30}_{-32}$, and $\kappa_\PQc=0.0^{+2.3}_{-2.8}$. This rules out the hypothesis that up- or down-type quarks in the first or second generation have  the same Yukawa couplings as those in the third generation, with a \CL greater than 95\%.
}

\hypersetup{
pdfauthor={CMS Collaboration},
pdftitle={Search for gamma H production and constraints on the Yukawa couplings of light quarks to the Higgs boson},
pdfsubject={CMS},
pdfkeywords={CMS, EFT, Higgs, anomalous couplings}}

\maketitle 

\section{Introduction}\label{sec:Introduction}

The observation of a Higgs boson ($\PH$) with a mass of 125\GeV by the ATLAS and CMS
Collaborations~\cite{Aad:2012tfa,Chatrchyan:2012xdj,Chatrchyan:2013lba} is consistent with
the expectations from the standard model (SM) of particle physics~\cite{StandardModel67_1, Englert:1964et,
Higgs:1964ia,Higgs:1964pj,Guralnik:1964eu,StandardModel67_2,StandardModel67_3}.
Constraints on \Hboson spin-parity properties and anomalous couplings
to electroweak gauge bosons $\PV=\PGg, \PZ, \PW$ (\HVV couplings) have been set by the CMS~\cite{Chatrchyan:2012jja,Chatrchyan:2013mxa,Khachatryan:2014kca,Khachatryan:2015mma,Khachatryan:2016tnr,
Sirunyan:2017tqd,Sirunyan:2019twz,Sirunyan:2019nbs,CMS:2021nnc,CMS:2022ley,CMS:2022uox}
and ATLAS~\cite{Aad:2013xqa,Aad:2015mxa,Aad:2016nal,Aaboud:2017oem,Aaboud:2017vzb,Aaboud:2018xdt,
ATLAS:2020rej,ATLAS:2022tan,ATLAS:2022tan,ATLAS:2023mqy} experiments.
The quantum numbers are found to be consistent with $J^{PC}=0^{++}$, 
but small anomalous \HVV couplings are still allowed. 
Such couplings are a subject of effective field theory (EFT) framework~\cite{deFlorian:2016spz}. 

\begin{figure*}[bht!]
\centering
\includegraphics[width=0.60\textwidth]{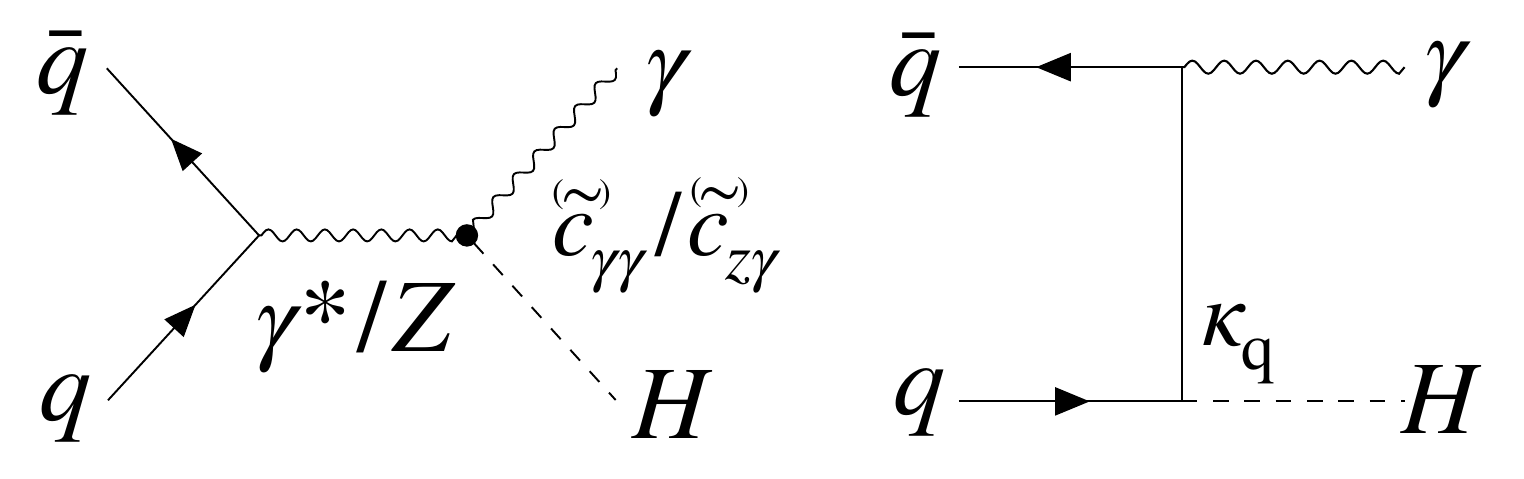}
\caption
{
Examples of Feynman diagrams illustrating \gammaH production at the LHC, highlighting the relevant couplings as discussed in the text.
On the left, production occurs via an effective loop-generated  \Hyy or \HZy interaction, denoted by a dot, 
which is the primary focus of this paper. On the right, production proceeds through \Hboson production
in \qqbar annihilation with photon radiation.
\label{fig:Feyn1}
}
\end{figure*}

The \HVV coupling measurements at the LHC have been performed using vector boson fusion (VBF), \ZH, and $\PW\PH$ production, 
and \HtoVV decays~\cite{Chatrchyan:2012jja,Chatrchyan:2013mxa,Khachatryan:2014kca,Khachatryan:2015mma,Khachatryan:2016tnr,
Sirunyan:2017tqd,Sirunyan:2019twz,Sirunyan:2019nbs,CMS:2021nnc,CMS:2022ley,CMS:2022uox,
Aad:2013xqa,Aad:2015mxa,Aad:2016nal,Aaboud:2017oem,Aaboud:2017vzb,Aaboud:2018xdt,
ATLAS:2020rej,ATLAS:2022tan,ATLAS:2022tan,ATLAS:2023mqy,ATLAS:2023yqk}. 
However, the production of \gammaH with an energetic photon recoiling against the \Hboson, 
similar to \ZH production as shown in Fig.~\ref{fig:Feyn1} (left), has not been directly studied at the LHC.
A recent search for $\PW\PW\PGg$  production~\cite{CMS:2023rcv} was interpreted as a search for
\gammaH, where a relatively soft photon is produced as a radiative correction to quark-antiquark annihilation $\qqbar\to\PH$, 
as shown in Fig.~\ref{fig:Feyn1} (right).
Searches for heavy resonances decaying into a photon and a hadronically decaying \Hboson have been performed at the 
LHC~\cite{CMS:2018cno,ATLAS:2018sxj,ATLAS:2020jeb}. 
Although the final-state topology in these searches resembles that of the processes of interest here, 
a direct interpretation within the EFT framework is not possible due to significant differences in the kinematic 
distributions, especially the photon energy spectrum, compared to those observed in the processes illustrated 
in Fig.~\ref{fig:Feyn1}. 

\begin{figure}[th!]
\centering
\includegraphics[width=0.49\textwidth]{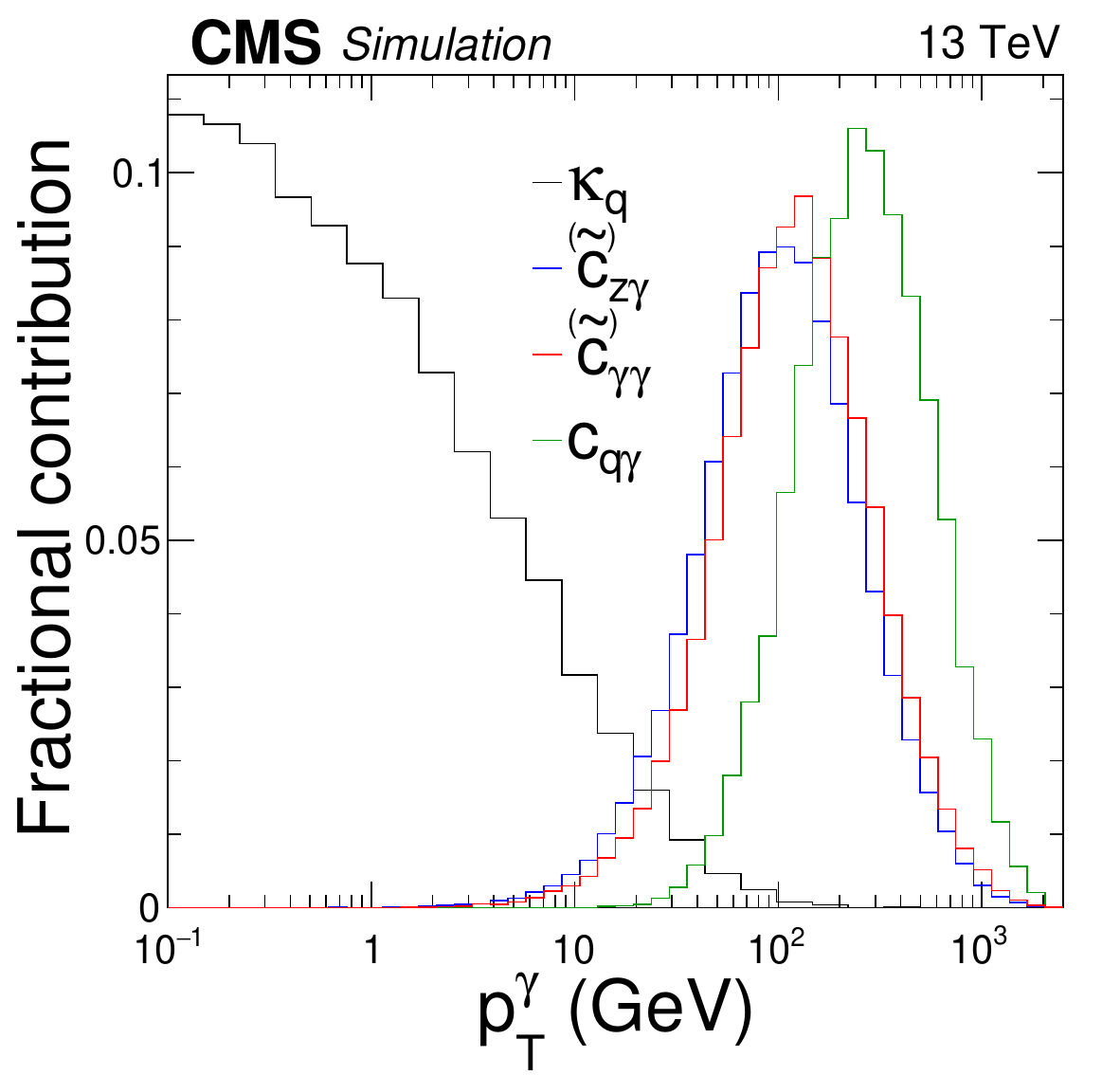}
\caption
{
The spectrum of the photon transverse momentum in \gammaH production, 
as generated by the leading-order diagrams shown in Figs.~\ref{fig:Feyn1} and~\ref{fig:Feyn2}. 
The four distributions correspond to production resulting from couplings
$\kappa_\PQq$, \czy (\tczy), \cyy (\tcyy), and $c_{\PQq\PGg}$. 
\label{fig:pT}
}
\end{figure}

The SM cross section of the associated photon and \Hboson production, \xsecGammaH, is expected to be less than 5\unit{fb}
at the LHC~\cite{Abbasabadi:1997zr} and is beyond the current experimental reach. 
However, new anomalous interactions may enhance such production,
as discussed in Refs.~\cite{Khanpour:2017inb,Shi:2018lqf,Davis:2021tiv}.
The main production mechanisms are shown in Fig.~\ref{fig:Feyn1} and the corresponding
transverse momentum (\pt) spectra of the associated photons
are shown in Fig.~\ref{fig:pT}~\cite{Davis:2021tiv, Brivio:SMEFTv1, Brivio:SMEFTv3, Alwall_2014}.
The latter are important for assessing the feasibility of detecting such final states.
In the case of bremsstrahlung radiation shown in Fig.~\ref{fig:Feyn1} (right), the photon \pt spectrum is soft,
a characteristic feature of quantum electrodynamics known as infrared divergence. 
In contrast, for Higgsstrahlung radiation shown in Fig.~\ref{fig:Feyn1} (left), the coupling is induced by a loop-generated 
higher-dimensional operator that includes a momentum factor in the numerator. 
This factor cancels the divergent term in the denominator, leading to an energetic photon recoiling against the \Hboson.
In this case, the photon \pt distribution peaks above 50\GeV and extends up to 
several hundred \GeVns, as shown in Fig.~\ref{fig:pT}~\cite{Davis:2021tiv}. 

The primary production mechanism targeted in this analysis is generated by the effective \HZy and \Hyy vertices, 
as shown in Fig.~\ref{fig:Feyn1} (left). 
The hard \pt spectrum resulting from anomalous \Hyy or \HZy interactions makes it possible to distinguish such final states.
These interactions could be generated by heavy particles in the loops leading
to both \CP-even and \CP-odd EFT operators, with Wilson coefficients \cyy, \czy, \tcyy, and \tczy
when expressed in the mass-eigenstate basis~\cite{deFlorian:2016spz}. 
Each of these operators can be represented as a combination of three operators in the weak-eigenstate 
basis~\cite{deFlorian:2016spz}. Overall, there are six operators to consider in the weak-eigenstate basis, 
with the Wilson coefficients $C^{\varphi W}$, $C^{\varphi\mathrm{B}}$, $C^{\varphi\mathrm{WB}}$, 
$C^{\varphi\mathrm{\widetilde{W}}}$, $C^{\varphi\mathrm{\widetilde{B}}}$, and $C^{\varphi\mathrm{\widetilde{W}B}}$.
Because of the smaller number of operators, the mass-eigenstate basis is chosen for this result, 
which can then be translated into the weak-eigenstate basis.

\begin{figure}[th!]
\centering
\includegraphics[width=0.49\textwidth]{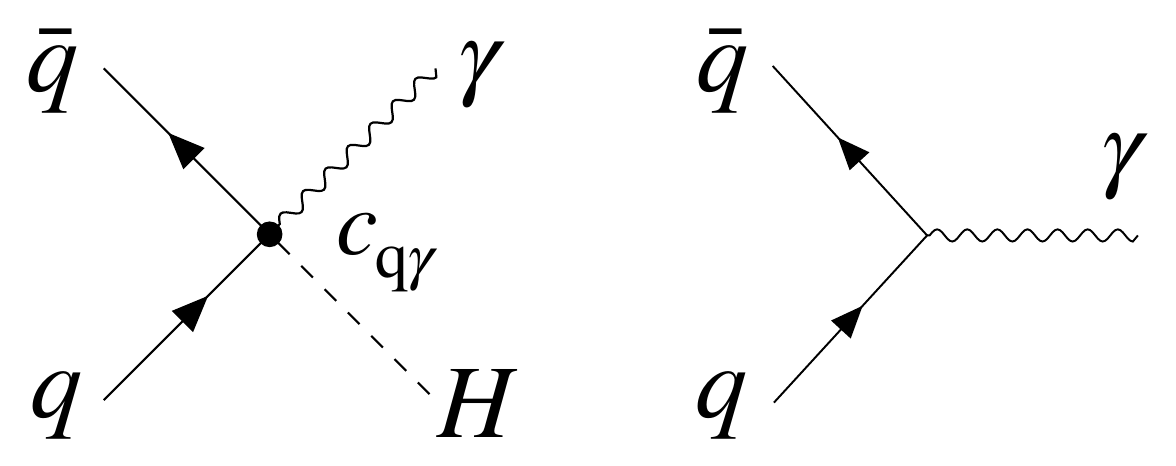}
\caption
{
Feynman diagrams describing the \qqbar annihilation with production of \gammaH 
through a point-like EFT operator (left) and with photon production (right). 
\label{fig:Feyn2}
}
\end{figure}

The $\PQq\PQq\PH\PGg$ point interaction, shown in Fig.~\ref{fig:Feyn2} (left),
with the linear combination of Wilson coefficients \CqB 
and \CqW, equivalent to one coefficient $c_{\PQq\PGg}$ in the mass eigenstate notation, 
can generate \gammaH production in \qqbar annihilation with a photon $\pt$ spectrum 
even harder than that produced by anomalous $\PH\PV\PGg$ couplings, 
as shown in Fig.~\ref{fig:pT}~\cite{Brivio:SMEFTv1, Brivio:SMEFTv3,Alwall_2014}.
The point interaction appears in the Feynman rules of the Standard Model Effective Field Theory (SMEFT)~\cite{Dedes:2017zog}.
However, the same beyond-the-SM (BSM) operators also contribute to the $\qqbar\PGg$ interaction, 
shown in Fig.~\ref{fig:Feyn2} (right), and $\qqbar\PZ$ interaction in the Feynman rules~\cite{Dedes:2017zog}. 
These operators are expected to be much better constrained by the more abundant processes without the presence of the \Hboson, such as Drell--Yan.
Although a fully optimized analysis of the Drell--Yan process to target various Wilson coefficients has not yet been performed at the LHC, 
current theoretical interpretations of published $\Pp\Pp \to\Pell\PGn$ and $\Pp\Pp \to\Pell\Pell$ results at the LHC indicate that the current 
constraints on the \CqB and \CqW Wilson coefficients from Drell--Yan are significantly stronger than those from processes 
involving the \Hboson~\cite{Theory:DrellYan}.
For these reasons, we do not consider this production diagram in this analysis. 

The Yukawa couplings of quarks $y_{\PQq}$ are incorporated into the effective Lagrangian 
that describes their interaction with the \Hboson as
\begin{equation}
\label{eq:LagrHqq}
{\cal L}(\PH\qqbar)=-y_{\PQq}\overline{\psi}_{\PQq}{\psi}_{\PQq}\PH,
\end{equation}
where $\overline{\psi}_{\PQq}$ and ${\psi}_{\PQq}$ denote the Dirac spinors for a quark $\PQq$,
which can be $\PQu$, $\PQd$, $\PQs$, $\PQc$, $\PQb$, or $\PQt$.
In the SM, $y_{\PQq}^\mathrm{\,SM}=m_{\PQq}/v$, with $v=246\GeV$ being the vacuum expectation value of the Higgs field. 
The quark Yukawa couplings may also be expressed in two alternative forms. 
First, we define $\kappaq=y_{\PQq}v/m_{\PQq}$, which is useful because $\kappaq^\mathrm{\,SM}=1$. 
Hence, $\kappaq$ may also be interpreted as the modifier for the coupling strength of the quark $\PQq$ to the \Hboson.
Second, we define $\barkappaq=y_{\PQq}v/m_{\PQb}$~\cite{Kagan:2014ila}, which is particularly useful
for comparing the hierarchy of the Yukawa couplings of light quarks, 
with respect to $\barkappab^\mathrm{\,SM}=1$ for the down-type quarks 
or $\barkappat^\mathrm{\,SM}=m_{\PQt}/m_{\PQb}$ for the up-type quarks. 
In these calculations, the quark masses are evaluated at the scale $\mu=125\GeV$, which are used to relate 
$\barkappaq=\kappaq m_{\PQq}/m_{\PQb}$. 

The emission of the \Hboson and a photon from a quark in the \qqbar annihilation, depicted in Fig.~\ref{fig:Feyn1} (right),
is the dominant \gammaH production channel in the SM~\cite{Abbasabadi:1997zr}.
However, this production mechanism would result in a photon with very low \pt, of less than 1\GeV on average, 
as determined from a dedicated simulation shown in Fig.~\ref{fig:pT}~\cite{Brivio:SMEFTv1, Brivio:SMEFTv3,Alwall_2014}.
Differentiating this production mechanism from others, such as gluon fusion (\ggH), becomes exceedingly challenging
because of the presence of soft photons, whether genuine or spurious, throughout the rest of the proton-proton ($\Pp\Pp$) collision event.
Nevertheless, an increased Yukawa coupling of light quarks could change the production rate associated with this mechanism.

\begin{figure}[t!]
\centering
\includegraphics[width=\cmsFigWidthF]{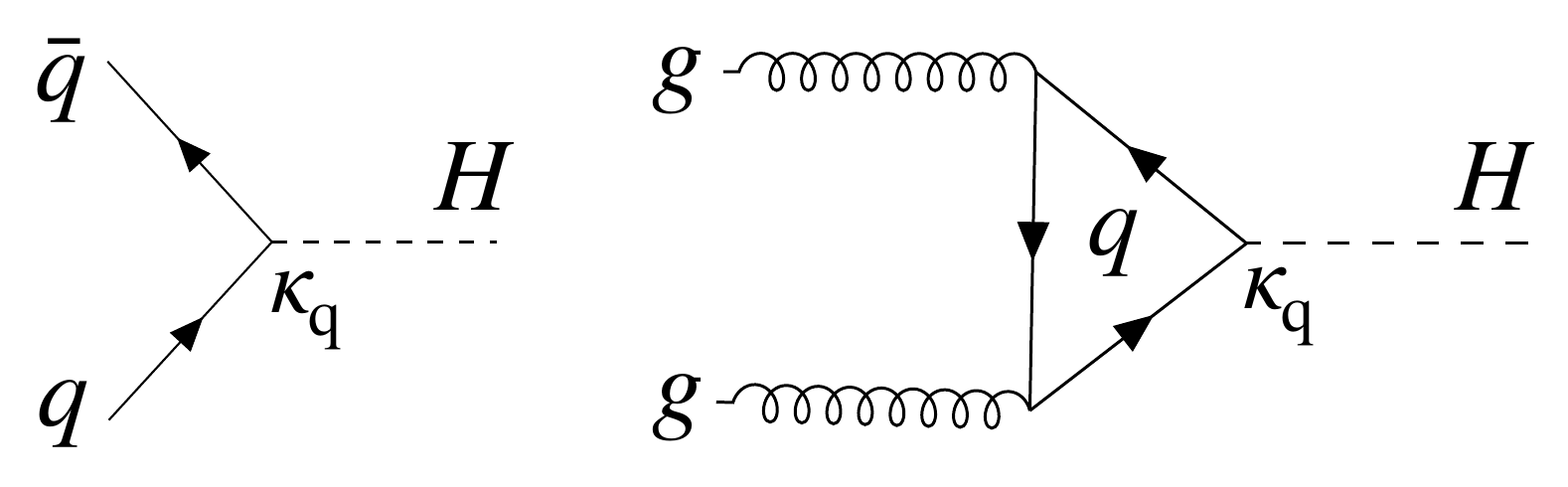}
\caption
{
Feynman diagrams describing the \Hboson production at LHC
through direct \qqbar annihilation (left) and gluon fusion production (right). 
\label{fig:Feyn3}
}
\end{figure}

This motivated an approach to constrain the Yukawa coupling of light quarks by imposing constraints on the \gammaH 
production rate~\cite{Aguilar-Saavedra:2020rgo}. 
However, the presence of an associated photon in this context does not effectively aid such an analysis due to the soft \pt spectrum. 
This photon merely represents a radiative correction to the direct \qqbar annihilation illustrated in Fig.~\ref{fig:Feyn3}~(left).
Therefore, we investigate the inclusive production of the \Hboson to explore the potential enhancement of the \Hqq coupling. 
This enhancement would also change the rate of the gluon fusion process depicted in Fig.~\ref{fig:Feyn3} (right), 
necessitating its consideration in this analysis as well.
The concept for this analysis was initially introduced in Refs.~\cite{Zhou:2015wra,Zhou:2019} 
and has been revisited more recently in Ref.~\cite{Balzani:2023jas}.

Other strategies have been proposed to constrain the Yukawa couplings of bottom and charm quarks.
Constraints on the $\PH\PQc\PQc$ coupling have been directly derived from searches for \Hboson 
decays into charm quarks~\cite{CMS:2022psv,ATLAS:2022ers}. 
Measurements of transverse momentum distributions in \Hboson production~\cite{Bishara:2016jga} have been used 
to derive experimental constraints, as outlined in Refs.~\cite{CMS:2018gwt,ATLAS:2022qef,CMS:2023gjz}.
A proposal was made to constrain the charm Yukawa coupling through a global fit to the \Hboson signal 
strengths~\cite{Perez:2015aoa}, although this approach overlooks the fact that SM assumptions are applied in these fits.
Radiative decays of the \Hboson to mesons could also be used to constrain the quark couplings~\cite{Bodwin:2013gca}.
However, there are currently no stringent experimental constraints on the couplings involving the $\PQs$, $\PQd$, and $\PQu$ quarks.
Nevertheless, it has been proposed that the transverse momentum and rapidity distributions of the \Hboson 
could help establish these constraints~\cite{Soreq:2016rae}.

This paper is organized as follows. 
Section~\ref{sec:Detector} provides a discussion of the CMS experiment and event reconstruction. 
Section~\ref{sec:Simulation} describes the modeling of signal and background processes. 
The selection of both \HfourL and \bbbar events is presented in Section~\ref{sec:EventSelection}.
Background estimation is discussed in Section~\ref{sec:BackgroundEstimation},
while the systematic uncertainties in the experimental measurement of cross sections are outlined in Section~\ref{sec:Systematics}.
This is followed by a description of the two analyses in Sections~\ref{sec:Results} and~\ref{sec:HqqResults}, with a summary of both analyses provided in Section~\ref{sec:Summary}.

For the analysis described in Section~\ref{sec:Results},
we follow the formalism and simulation tools from Ref.~\cite{Davis:2021tiv} and set
upper limits on the \gammaH production cross section, along with
simultaneous constraints on four anomalous \HZy and \Hyy
couplings in the mass-eigenstate EFT basis. 
The analysis is based on the signature of an \Hboson boosted in the direction transverse to the beam and
recoiling against a high-energy photon. 
Two final states in the decay of the \Hboson are considered: \Htobbbar, with a branching fraction of $\mathcal{B}_{\bbbar}=0.577$, 
and \Hell, with $\mathcal{B}_{4\Pell}=0.000128$, both evaluated at $m_{\PH}=125.38\GeV$~\cite{deFlorian:2016spz}, 
where $\Pell$ refers to either an electron or a muon.
The boosted topology of the \Hboson gives preference to the former channel because of its much larger~$\mathcal{B}$,
but both channels are analyzed for completeness. 

For the analysis in Section~\ref{sec:HqqResults}, we investigate the rate of \Hboson production 
for potential enhancements of the Yukawa couplings between light quarks and the \Hboson. 
This includes examining modifications to both direct quark-antiquark annihilation and gluon fusion loop processes. 
The enhanced \Hqq couplings would lead to an increased width of the \Hboson and reduced rate of the
on-shell process despite the increase of the amplitudes for the quark-antiquark annihilation and gluon fusion loop processes. 
The absence of significant changes in the SM rate imposes restrictions on the Yukawa couplings of light quarks.
Some assumptions are required for this analysis. 
For instance, we assume that the Yukawa couplings for the bottom and top quarks take their SM values, 
and that the \HVV coupling strength does not exceed its SM value.
This analysis is conducted in the four-lepton decay channel because of its high purity, which enables inclusive reconstruction 
regardless of the \Hboson's production mechanism and kinematic properties.

Tabulated results are provided in the HEPData record for this analysis~\cite{hepdata}.

\section{The CMS detector and event reconstruction}
\label{sec:Detector}
The central feature of the CMS apparatus is a superconducting solenoid of 6\unit{m} internal diameter, 
providing a magnetic field of 3.8\unit{T}. Within the solenoid volume are a silicon pixel and strip tracker, 
a lead tungstate crystal electromagnetic calorimeter (ECAL), and a brass and scintillator hadron calorimeter (HCAL), 
each composed of a barrel and two endcap sections. Forward calorimeters extend the pseudorapidity ($\eta$)
coverage provided by the barrel and endcap detectors. Muons are measured in gas-ionization detectors 
embedded in the steel flux-return yoke outside the solenoid. A more detailed description of the CMS detector, 
together with a definition of the coordinate system used and the relevant kinematic variables, can be found 
in Ref.~\cite{Chatrchyan:2008zzk,CMSTrackerGroup:2020edz}. 

The data from the CMS experiment at the LHC were recorded between 2016 and 2018, corresponding 
to an integrated luminosity of 138\unit{\fbinv} at a proton-proton center-of-mass collision energy of 13\TeV.
Events of interest were selected using a two-tier trigger system.
The first level, composed of custom hardware processors, uses information from the calorimeters and muon 
detectors to select events at a rate of around 100\unit{kHz} within a fixed latency of about 4\mus~\cite{CMS:2020cmk}. 
The second level, known as the high-level trigger, consists of a farm of processors running a version of the 
full event reconstruction software optimized for fast processing, and reduces the event rate to around 1\unit{kHz} 
before data storage~\cite{CMS:2016ngn,Hayrapetyan_2024}.

The primary vertex is taken to be the vertex corresponding to the hardest scattering in the event, 
evaluated using tracking information alone, as described in Section 9.4.1 of Ref.~\cite{CMS-TDR-15-02}. 

A particle-flow algorithm~\cite{Sirunyan:2017ulk} aims to reconstruct and identify each individual particle 
in an event, with an optimized combination of information from the various elements of the CMS detector. 
The energy of photons is obtained from the ECAL measurement. The energy of electrons is determined 
from a combination of the electron momentum at the primary interaction vertex as determined by the tracker, 
the energy of the corresponding ECAL cluster, and the energy sum of all bremsstrahlung photons spatially 
compatible with originating from the electron track. The energy of muons is obtained from the curvature of 
the corresponding track. The energy of charged hadrons is determined from a combination of their momentum 
measured in the tracker and the matching ECAL and HCAL energy deposits, corrected for the response 
function of the calorimeters to hadronic showers. Finally, the energy of neutral hadrons is obtained from 
the corresponding corrected ECAL and HCAL energies.

Jets are clustered from particle-flow candidates using the anti-\kt algorithm~\cite{Cacciari:2008gp,Cacciari:2011ma} 
with a distance parameter of either 0.4 (AK4 jets) or 0.8 (AK8 jets). The jet momentum is defined as the vector 
sum of all particle momenta in a jet and is found from simulation to be, on average, within 5--10\% of the true 
momentum over the entire \pt spectrum and detector acceptance~\cite{CMS:2016lmd}.

\section{Signal and background simulation}
\label{sec:Simulation}

Simulated samples of signal and background events are produced using various Monte Carlo (MC) event generators, 
with the CMS detector response modeled by \GEANTfour~\cite{Agostinelli2003250}. The effects of pileup are modeled 
assuming a total inelastic $\Pp\Pp$ cross section of {69.2\unit{mb}}~\cite{CMSLHCppXsec}. All simulated event samples 
are weighted to match the distribution of the expected pileup profile of the data.
The parton distribution functions (PDFs) used in this paper belong to the Neural Network Parton Distribution Functions (NNPDF) sets, versions 3.0 and 3.1~\cite{Ball:2011uy}.
All MC samples are interfaced with \PYTHIA~8.230~\cite{Sjostrand:2014zea} for parton showering.
The $\PH\to \PQb\PAQb$ decay is also simulated with \PYTHIA.

The \jhugen~7.5.2~\cite{Gao:2010qx,Bolognesi:2012mm,Anderson:2013afp,Gritsan:2016hjl,Gritsan:2020pib} 
MC program is used to simulate the \gammaH production at leading order (LO) in quantum chromodynamics (QCD) and electroweak interactions.
The same program is used to simulate the \HfourL decay in all MC samples involving this final state. 
The \mela~\cite{Gao:2010qx,Bolognesi:2012mm,Anderson:2013afp,Gritsan:2016hjl,Gritsan:2020pib} 
package, which contains a library of matrix elements from \jhugen for the \Hboson signal and \MCFM 7.0.1~\cite{MCFM} 
for the $4\Pell$ background, is used for MC sample reweighting, and for optimal discriminant calculations. 
The \gammaH samples are generated for various values of 
\cyy, \czy, \tcyy, and \tczy, 
and are reweighted to any hypothesis of interest. 
The production of the \Hboson through gluon fusion, \VBF, in association with a $\PZ$ or $\PW$ boson, 
or with a \ttbar pair is simulated using \POWHEG~v2~\cite{Frixione:2007vw,Bagnaschi:2011tu,Nason:2009ai,Luisoni:2013kna,Hartanto:2015uka}
at next-to-leading order (NLO) in QCD. Production in association with a \bbbar pair or single top quark is simulated only at LO in QCD via \jhugen.
All SM cross sections of the \Hboson production processes are matched to the recommendation in Ref.~\cite{deFlorian:2016spz}. 
There are no final-state photons in any of the SM processes for \Hboson production simulated at the matrix-element level.
Therefore, the reconstructed photons in the SM simulation originate either from genuine photons produced during parton showering and hadronization, 
or from other SM particles that satisfy the photon selection criteria.

The background in the \bbbar channel mostly consists of nonresonant QCD multijet production with a real or misidentified photon, 
as well as resonant \Zgamma production. 
The \bbbar channel also makes use of a control event category that vetoes the presence of high-energy photons for which 
the QCD multijet, \Wjets, and \Zjets processes are the significant contributions. 
The \Wgamma and \Zgamma processes are modeled at NLO, while the rest are modeled at LO. 
All processes are generated using \MGvATNLO~v2.4.2~\cite{Alwall_2014}. 

The vector boson samples include boson decays to all flavors of quarks as well as up to three (four) extra partons 
at the matrix element level for \Wjets (\Zjets) and up to one extra parton for \Vgamma samples.
Jets from the matrix element calculation and the parton shower description are matched using the MLM prescription~\cite{Alwall_2007} 
for \Vjets samples and \fxfx matching~\cite{Frederix_2012} for \Vgamma samples.
Correction factors are applied to the \Vjets samples to match the generator-level \pt 
distributions with those predicted by the highest available order in the perturbative expansion. 
The QCD NLO corrections are derived using \MGvATNLO, simulating \PW and \PZ boson production with up to two additional partons and \fxfx matching to the parton shower.
The electroweak NLO corrections are taken from theoretical calculations in Ref.~\cite{Lindert_2017}. Electroweak NLO corrections, taken from Ref.~\cite{Denner_2016}, are also applied to the \Zgamma sample.

In the $4\Pell$ channel, the main background, $\qqbar\to\ZZ/\PZ\PGg^*\to 4\Pell$, is estimated from simulation with \POWHEG at NLO in QCD.
A fully differential cross section has been computed at next-to-next-to leading order (NNLO) in QCD~\cite{Grazzini:2015hta}, but it is not yet available
in a parton-level event generator. Therefore the NNLO/NLO QCD correction is applied as a function of \mell.
The $\Pg\Pg\to\ZZ/\PZ\PGg^*\to 4\Pell$ background and electroweak background production of $4\Pell$ are generated
with the \jhugen package, which relies on the \MCFM matrix elements~\cite{Campbell:2011bn,Campbell:2013una,Campbell:2015vwa}. 
In order to include higher-order corrections, the $K$ factors calculated for the \Hboson signal~\cite{deFlorian:2016spz}
are applied to these two background processes. 

\section{Event selection}
\label{sec:EventSelection}

Two mutually exclusive selection requirements are used in the \HfourL and \bbbar channels. 
They are described in Sections~\ref{ss:EventSelectionZZ} and~\ref{ss:EventSelectionbb}, respectively.
In both channels, two categories of events are selected. 
In one case, an \Hboson candidate is selected to be associated with a high-energy photon,
and the category is called $\PGg$-tagged. 
The other selected events are assigned to the Untagged category. 

\subsection{Event selection in the four-lepton channel}\label{ss:EventSelectionZZ}

The selection of $4\Pell$ events (\PH boson candidates) and associated photons closely follows the methods used in the earlier
analyses~\cite{Sirunyan:2017exp,Sirunyan:2017tqd,Sirunyan:2019twz,CMS:2021ugl}.
The main triggers select either a pair of electrons or muons, or an electron and a muon.
The minimal \pt of the leading electron (muon) is 23\,(17)\GeV,
while that of the subleading lepton is 12\,(8)\GeV. To maximize the signal acceptance,
triggers requiring three leptons with lower \pt thresholds and no isolation requirement are also used,
as are isolated single-electron and single-muon triggers with thresholds of 27 and 22\GeV
in 2016, or 35 and 27\GeV in 2017 and 2018, respectively. The overall trigger efficiency for
simulated signal events that pass the full selection chain of this analysis is larger than 99\%.
The trigger efficiency is measured in data using a sample of $4\Pell$ events collected
by the single-lepton triggers and is found to be consistent with the expectation from simulation.

Electrons (muons) are reconstructed within the geometrical acceptance defined by a requirement
on the pseudorapidity $\abs{\eta} < 2.5\,(2.4)$ for $\PT > 7\,(5)\GeV$ with
an algorithm that combines information from the ECAL (muon system) and the tracker.
A dedicated algorithm is used to collect the final-state radiation off leptons~\cite{Sirunyan:2017exp}.
To distinguish leptons from prompt $\PZ$ boson decays from those originating in hadron decays within jets, 
a lepton isolation requirement is applied in the analysis~\cite{Sirunyan:2017exp}. In addition, to suppress non-prompt 
and misidentified leptons, primary leptons are required to satisfy $\abs{\text{SIP}} < 4$, 
where the significance of the impact parameter (SIP) is defined as $\text{SIP} = \text{IP}/\sigma_{\text{IP}}$.
Here, IP refers to the lepton's three-dimensional impact parameter relative to the selected primary vertex, and 
$\sigma_{\text{IP}}$  is its associated uncertainty.

\begin{figure}[!tbh]
    \centering
    \includegraphics[width=0.45\textwidth]{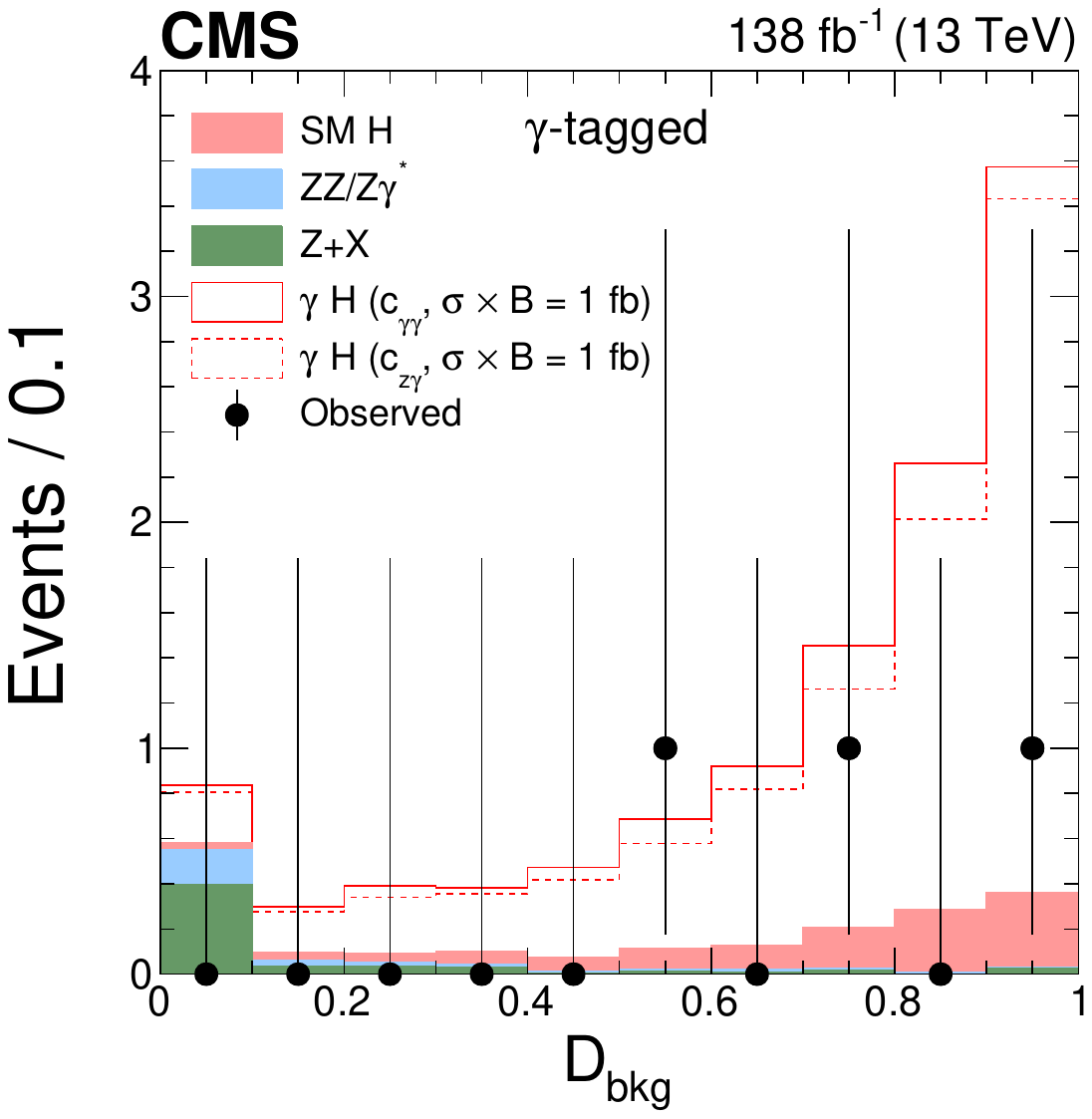}
    \includegraphics[width=0.45\textwidth]{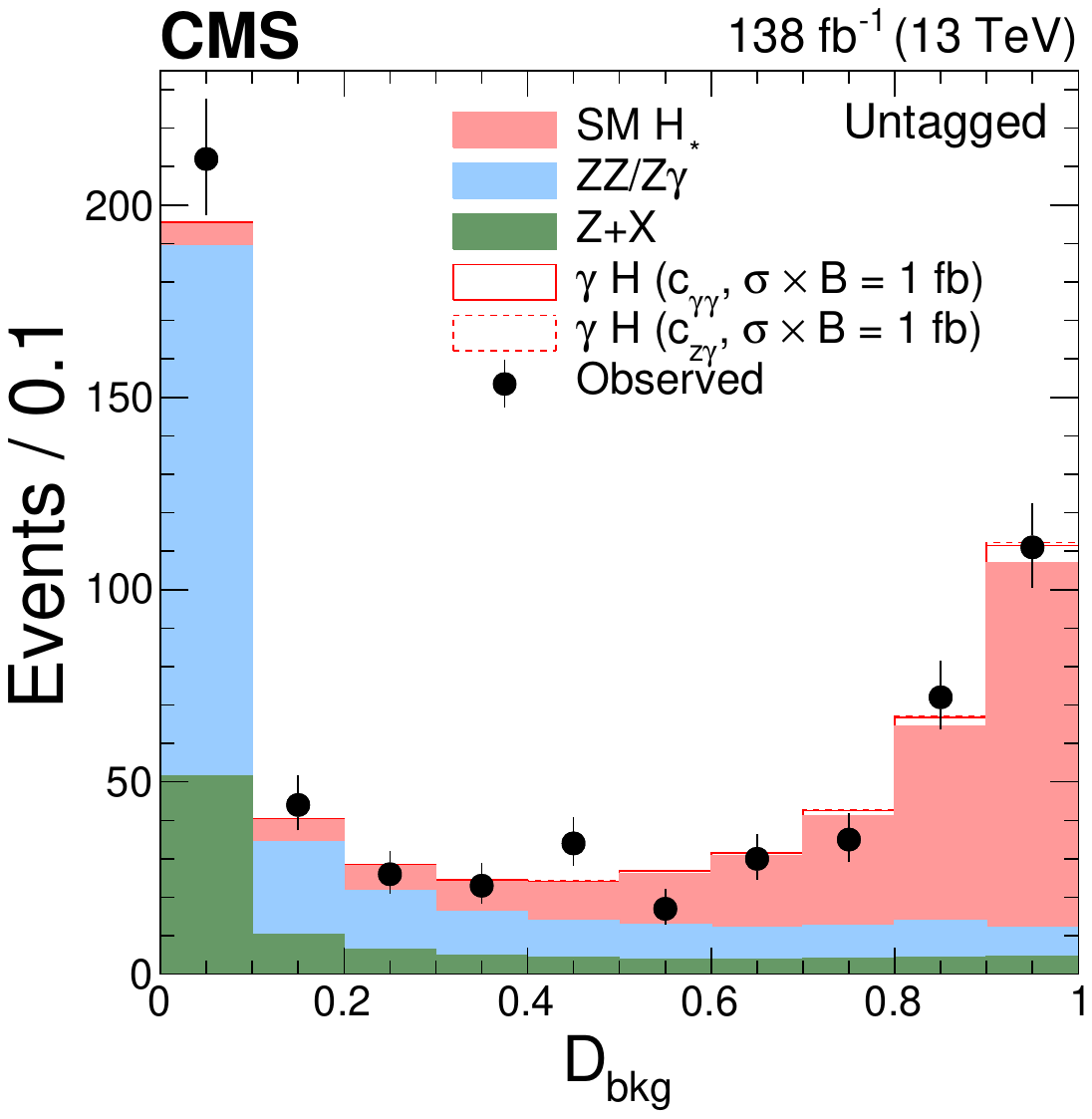}
    \caption{
    Distributions of events for the \Dbkg observable in the $\PGg$-tagged (\cmsLeft) and Untagged (\cmsRight) categories 
    of the \HfourL candidate events. 
    Observed events (black markers) and expected background estimates (solid histograms) from MC simulation (ZZ/Z$\PGg^*$) 
    or control samples in data (\ZX) are shown. 
    The \gammaH signal contribution, stacked on top of background, is shown with an open histogram for 
    an assumed cross section of $\xsecGammaH\,\mathcal{B}_{4\Pell}=1\unit{fb}$ 
    for either the \cyy (solid) and \czy (dashed) coupling hypothesis.}
    \label{fig:postfit_H4l}
\end{figure}

The $\PGg$-tagged category requires the presence of a photon passing loose cut-based identification requirements~\cite{CMSphotonPerf}
and $\pt > 150\GeV$. All $4\Pell$ events that pass the selection requirements 
but do not include a photon that passes the requirements for the $\PGg$-tagged category enter the Untagged category. 
We consider three mutually exclusive channels: $\PH\to4\Pe$, $4\Pgm$, and $2\Pe 2\Pgm$.
At least two leptons are required to have $\pt > 10\GeV$, and at least one is required to have $\pt > 20\GeV$.
All four pairs of oppositely charged leptons that can be built with the four leptons are required
to satisfy $m_{\Pellp\Pellm} > 4\GeV$ regardless of lepton flavor. The $\PZ$ boson candidates are required to satisfy 
the condition $12 < m_{\Pellp\Pellm} < 120\GeV$, where the invariant mass of at least one of the $\PZ$ boson candidates must
be larger than 40\GeV. The region between 105 and 140\GeV in the four-lepton invariant mass \mell
is considered in this analysis. 

Each event in the \HfourL channel is characterized by an optimal discriminant to separate signal from background. 
The discriminant is based on the \Hboson decay without using information related to the production mechanism, defined as
\begin{equation}
  \Dbkg\left(\boldsymbol{\Omega}\right) = \frac{\Psig\left(\boldsymbol{\Omega}\right) }
        {\mathcal{P}_\text{sig}\left(\boldsymbol{\Omega}\right) +\Pbkg\left(\boldsymbol{\Omega}\right) },
\label{eq:melaD}
\end{equation}
where $\Psig$ is the probability for the event to be consistent with the SM signal and $\Pbkg$ 
is the probability for the same event to be consistent with the dominant $\qqbar\to\ZZ/\PZ\PGg^*\to 4\Pell$ background.
The probabilities $\mathcal{P}$ are calculated from the matrix elements provided by the \mela package 
using kinematic observables $\Omega$, which are derived from the reconstructed momenta of the four leptons.
These probabilities are normalized to give the same integrated cross sections in the relevant phase space of each process.
Each matrix element probability is also multiplied by an empirical \mell parameterization, which includes resolution 
effects in the \mell distribution. 
The distributions of the \Dbkg discriminant in the two categories of events are shown in Fig.~\ref{fig:postfit_H4l}.

\subsection{Event selection in the \texorpdfstring{\bbbar}{bbbar} channel}\label{ss:EventSelectionbb}

In the \bbbar channel, the $\PGg$-tagged category aims to capture signal $\PGg \PH$ events, 
while the Untagged category targets events with a boosted $\PZ\to\bbbar$  and a recoiling jet. 
The Untagged category also includes $\PH\to\bbbar$ events, although their contribution is negligible.
The triggers in the $\PGg$-tagged category require events to have a photon with $\pt > 175(200)\GeV$ in 2016 (2017, 2018). 
The threshold is much lower than the \pt requirement placed in the offline selection described below. 
This makes the trigger selection almost fully efficient for the events that pass the offline selection.

The offline selection in the $\PGg$-tagged category requires that the leading \pt photon satisfies $\pt > 300\GeV$, $\lvert \eta \rvert < 2.4$, 
and the cut-based identification criteria at the tight working point~\cite{CMSphotonPerf}.
An \Hboson candidate is defined as the AK8 jet with the highest \pt that also satisfies $\Delta R$ $(\text{jet, leading photon})>0.8$, 
where $\Delta R(1,2) \equiv \sqrt[htb]{\smash[b]{(\eta_{1}-\eta_{2})^{2}+(\varphi_{1}-\varphi_{2})^{2}}}$ is the distance between two 
objects in the pseudorapidity--azimuthal-angle plane. The \Hboson candidate is required to have $\pt > 300\GeV$ and $\lvert \eta \rvert < 2.4$. 
The analysis uses a mass decorrelated regression algorithm based on the \textsc{ParticleNet}~\cite{PNet_HQu_LGouskos} graph convolutional 
neural network architecture to predict the jet mass $\mpnet$~\cite{CMS-DP-2021-017,CMS-DP-2022-005,PhysRevLett.131.041803}. 
The \Hboson candidates are required to have the regressed mass $\mpnet > 60\GeV$.
A veto is placed on the presence of leptons, making the selected set of events orthogonal to that in the $4\Pell$ channel. 
To veto an event, a muon is required to have $\pt > 20\GeV$, $\lvert \eta \rvert < 2.4$, and must pass the loose 
identification and isolation criteria~\cite{MuonsCMSRun2}.
Similarly, a veto electron is required to have $\pt > 20\GeV$, $\lvert \eta \rvert < 2.4$, and must pass
the cut-based identification at the veto working point~\cite{CMSphotonPerf}. 
Finally, a veto is placed on any b-tagged AK4 jet with $\pt > 30\GeV$ and $\Delta R(\text{jet, H candidate}) > 0.8$ in order 
to suppress the background arising from \ttbar production. 
The AK4 jets are identified as originating from $\cPqb$ quarks using the DeepJet algorithm~\cite{DeepJet}.
Events are classified based on how consistent the \Hboson candidate is with a boosted decay to $\bbbar$, using the \textsc{ParticleNet} algorithm, as described later in this section.

The Untagged category uses multiple triggers, based on energetic jet activity in the event, for the online selection. 
One trigger sets a requirement on the scalar \pt sum (\HT) of all AK4 jets. The \HT requirement is set to $>$800, $>$900, or $>$1050\GeV. A second trigger requires a presence of an AK8 jet with mass $>$30\GeV
and $\pt>360$, 400, or 420\GeV. The third trigger condition is the presence of an AK8 jet with $\pt>450$ or 500\GeV. 
The final trigger requires $\HT > 650$, 700, or 800\GeV together with an AK8 jet having a mass $>$50\GeV. 
The listed differing thresholds depend on the data-taking conditions with lower values typically being used in 2016. 
Events are selected if they pass any of the trigger conditions. 

In the offline selection of the Untagged category, a \PZ boson candidate is defined as the AK8 jet with the highest \pt. 
It is required to satisfy $\pt > 450\GeV$, $\mpnet > 60\GeV$, and $\lvert \eta \rvert < 2.4$. Conditions of $\pt>200\GeV$ and 
$\lvert \eta \rvert < 2.4$ are also applied on the subleading (recoil) AK8 jet in order to further suppress the QCD multijet background.
The Untagged category uses the same vetoes as the $\PGg$-tagged category. In addition, it places a veto on the 
presence of a photon with $\pt>300\GeV$, making it orthogonal to the $\PGg$-tagged category.

\begin{figure*}[htb]
\centerline{
      \includegraphics[width=0.32\textwidth]{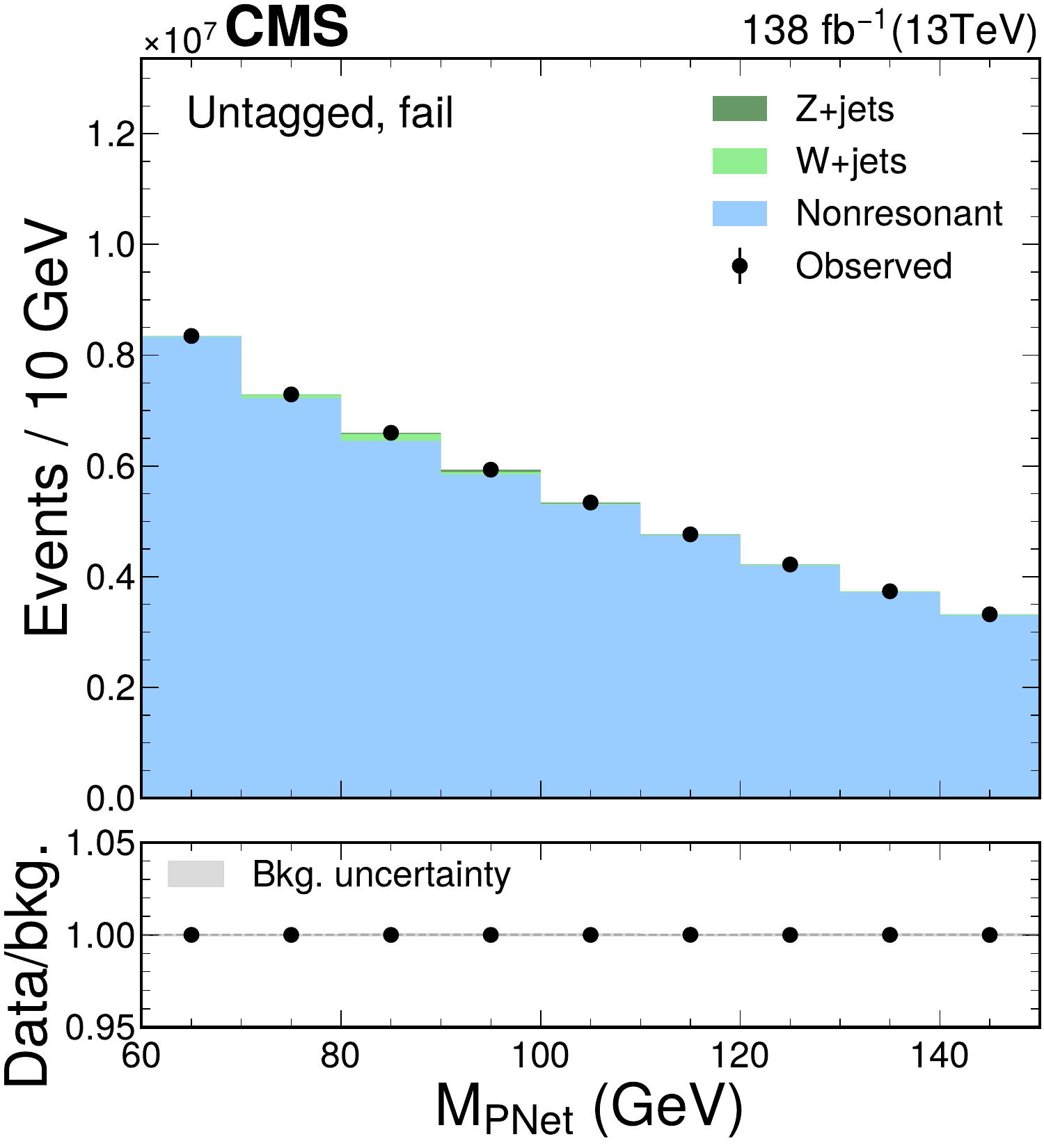}
      \includegraphics[width=0.32\textwidth]{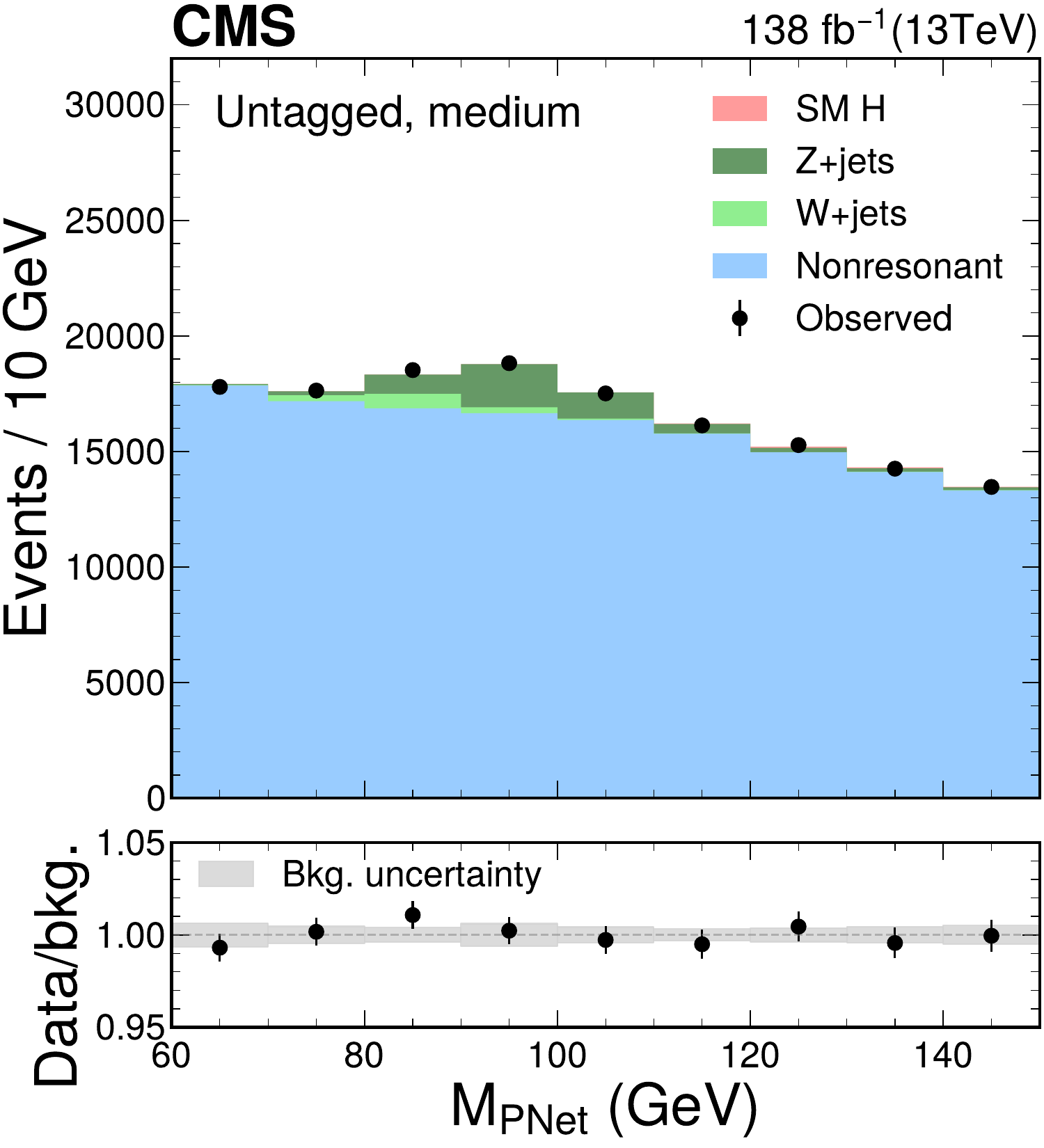}
      \includegraphics[width=0.32\textwidth]{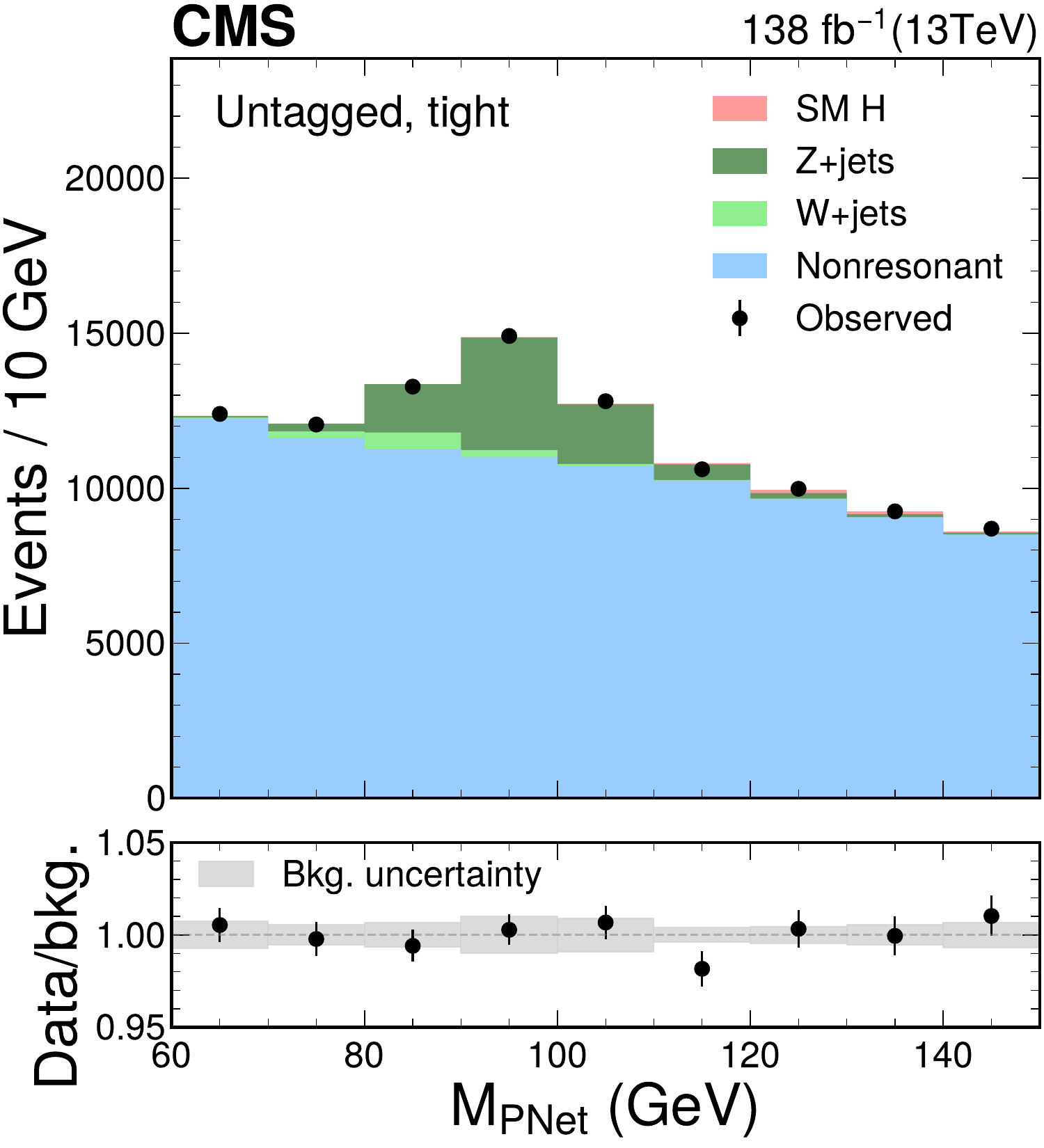} 
}
\centerline{
      \includegraphics[width=0.32\textwidth]{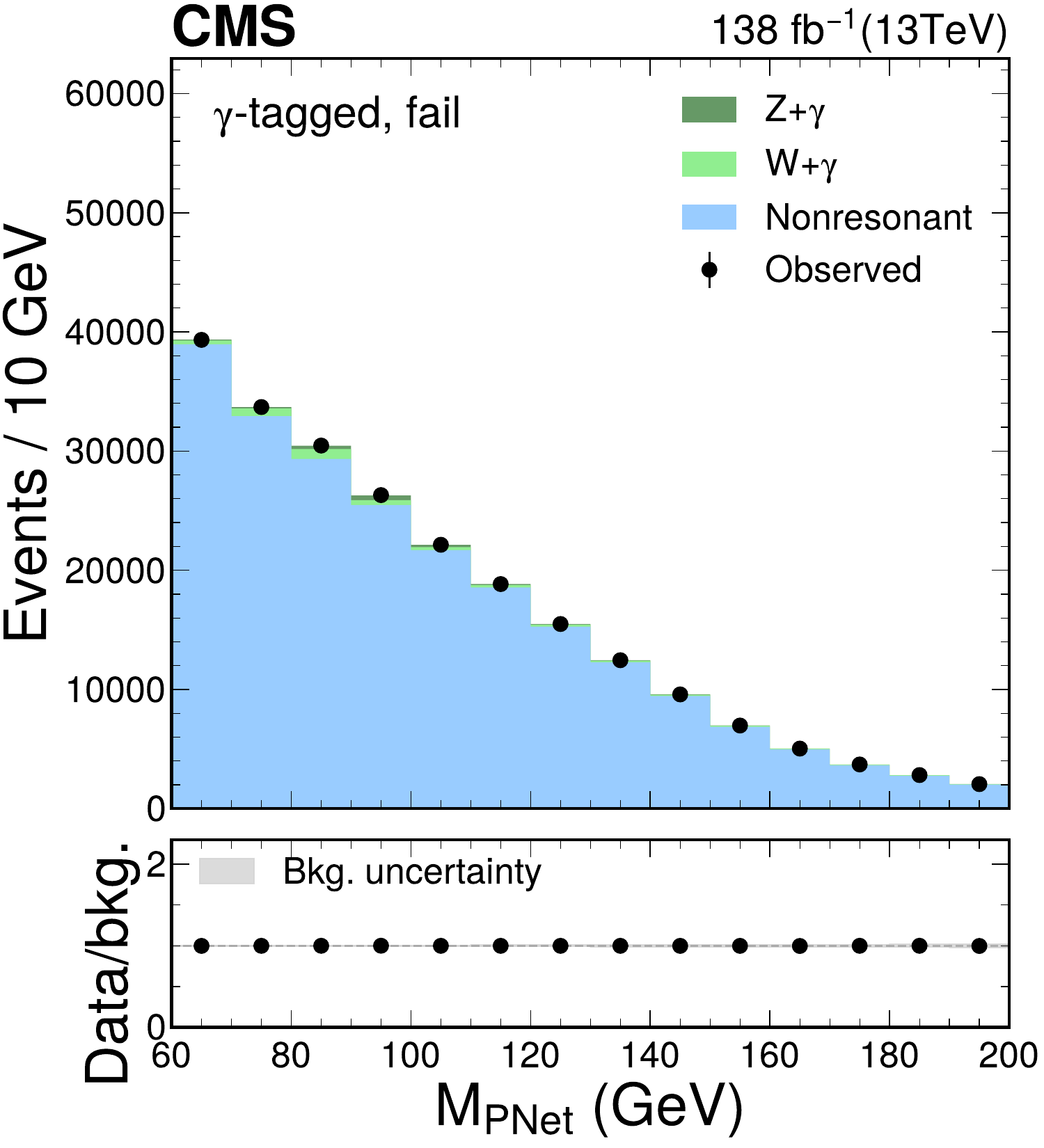}
      \includegraphics[width=0.32\textwidth]{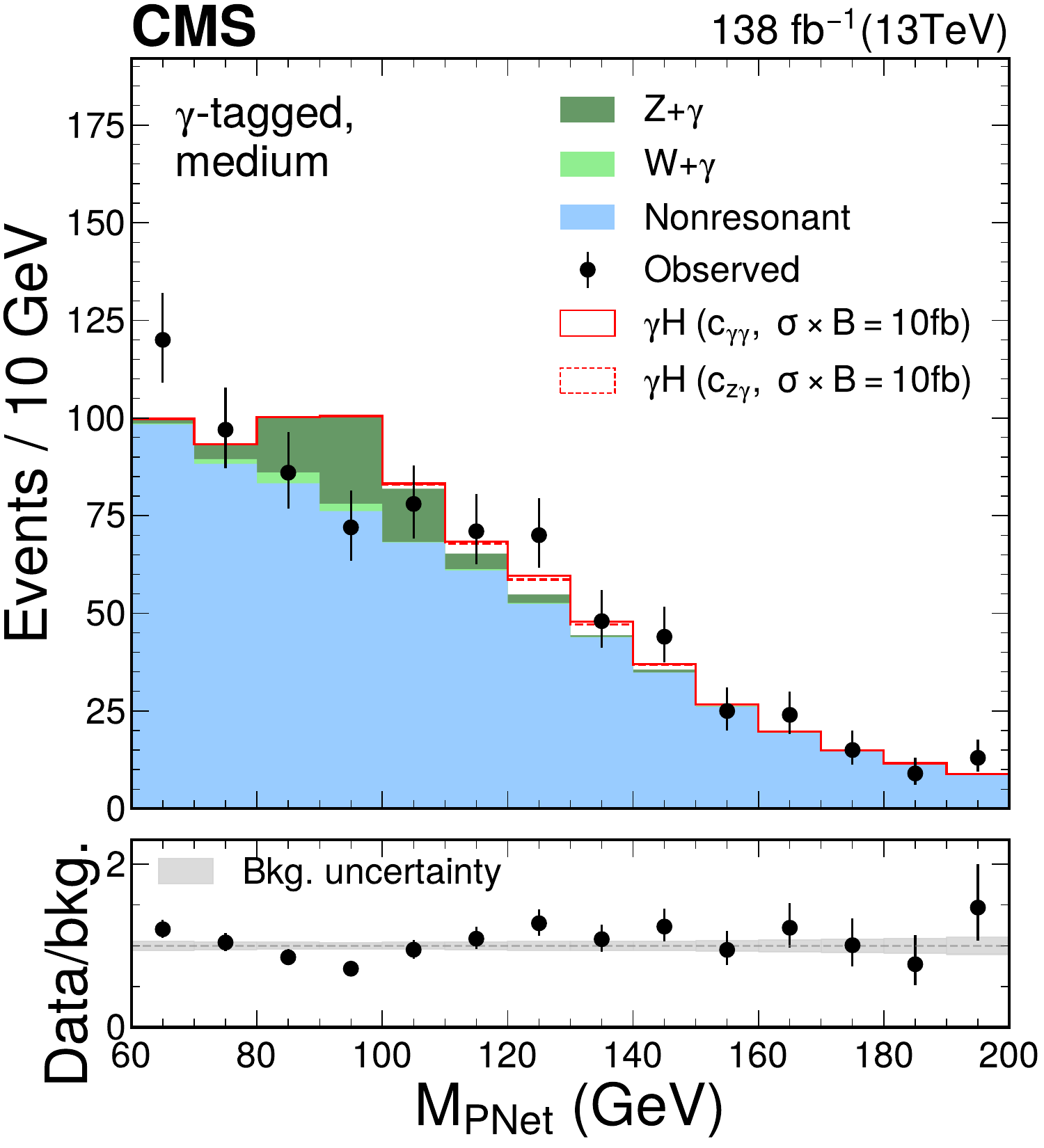}
      \includegraphics[width=0.32\textwidth]{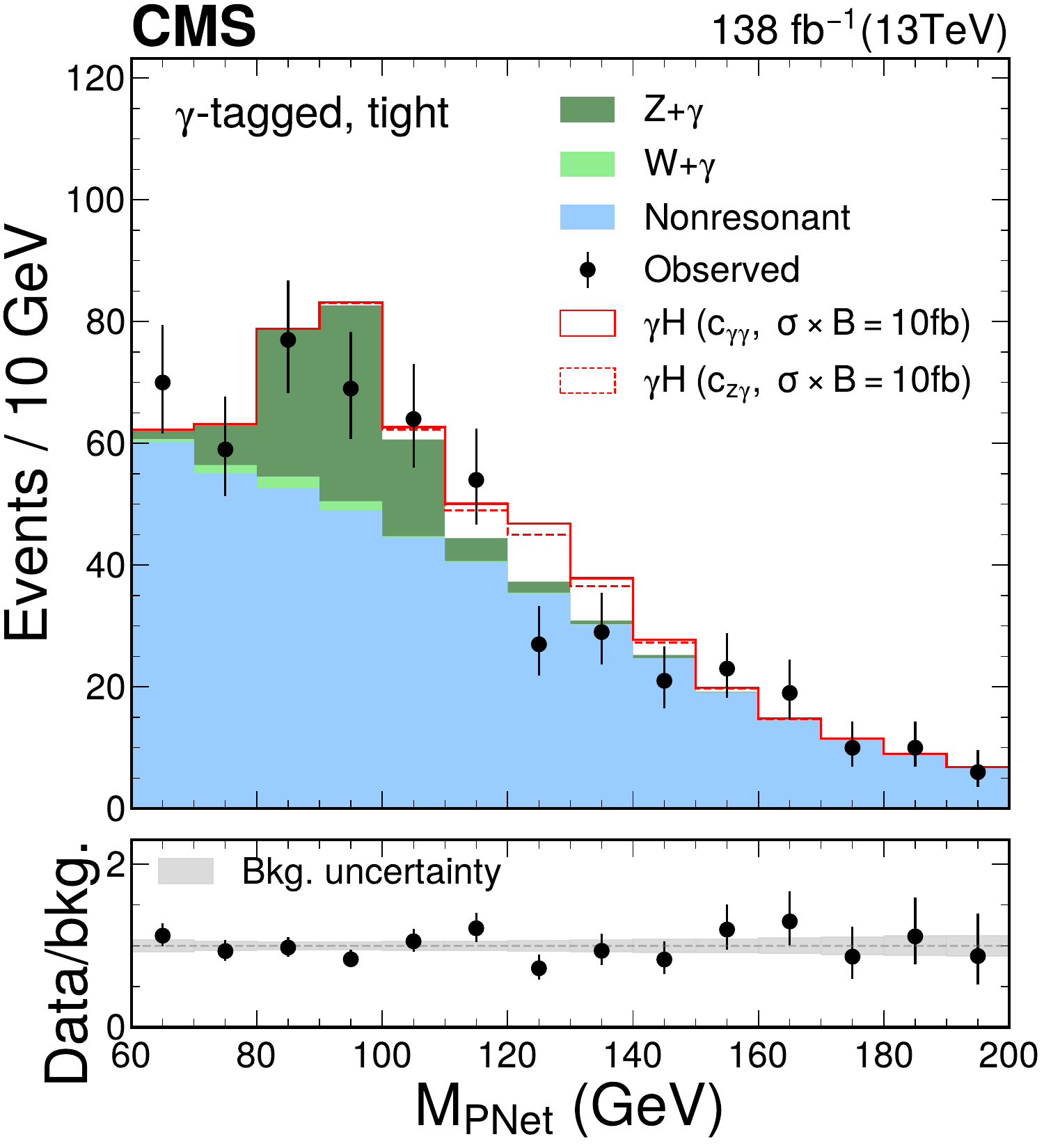}
      }
    \caption{The \mpnet distributions for the number of observed events (black markers) compared with 
    the backgrounds estimated in the fit to the data (filled histograms) in the \bbbar channel. Fail (left), 
    medium (middle) and tight (right) regions of the $\PGg$-tagged (lower) and Untagged (upper) categories 
    are shown. The signal contribution, stacked on top of background, is shown with an open histogram for 
    an assumed cross section of $\xsecGammaH\,\mathcal{B}_{\bbbar}=10\unit{fb}$.}
    \label{fig:postfit_bb}
\end{figure*}

The trigger efficiency of the Untagged event category is measured in data where online selection requires an isolated muon 
with $\pt>24\GeV$ ($27\GeV$ in 2017) or a muon with $\pt>50\GeV$ without any isolation requirements. The above offline 
selection is applied to these events in order to measure the trigger efficiency by counting the number of events satisfying the 
trigger selection. The trigger efficiency is expressed as a function of the \PZ boson \pt. It ranges from around 65\% at 450\GeV and 
becomes fully efficient at 600\GeV. Simulated events are weighted by this efficiency as a function of the \PZ boson \pt and \mpnet.

The \textsc{ParticleNet} architecture is also employed to discriminate the decays of a boosted massive particle \PX, which could be an 
\PH or a \PZ boson, to a pair of \PQb quarks against a background of other jets, using the properties of the jet constituents
as features. The multiclassifier \textsc{ParticleNet} algorithm~\cite{PNet_HQu_LGouskos} outputs several variables, each in the range 0--1, 
and each of which can be interpreted as the probability of a jet having originated from a certain decay, such as from a massive 
resonance \Xbb (\PXbb) or from a light-flavored quark or a gluon (\PQCD). In this analysis, the \textsc{ParticleNet} score 
is defined as $\PXbb /(\PXbb + \PQCD)$.

The \textsc{ParticleNet} scores of the \PH and \PZ jets are used to classify events into mutually exclusive regions using high- and medium-purity working points~\cite{CMS-PAS-BTV-22-001}. If the candidate jet passes the high-purity working point, the 
event is sorted into the ``tight'' (T) region. If the candidate jet passes the medium-purity working point and fails the high-purity working point, the event is sorted into the ``medium'' (M) region. All other events are sorted into the ``fail'' (F) region.

To increase the channel sensitivity, the $\PGg$-tagged category is further split into two regions based on the photon 
\pt, 300--400\GeV and $>$400\GeV. This gives six regions in total in the $\PGg$-tagged category and three 
regions in the Untagged category.
The distributions of the \mpnet observable, after fitting the background model to the data, are shown in Fig.~\ref{fig:postfit_bb} 
with the two \pt regions of the $\PGg$-tagged category combined. The background prediction mostly agrees with the data. 
The largest discrepancy is seen in the M region of the $\PGg$-tagged category where a deficit of events under the \PZ peak 
is observed. However, a goodness-of-fit test~\cite{CMS:2024onh} confirmed the agreement between the data and the estimated background 
with the p-value~\cite{pvalue} greater than 0.05, indicating no significant deviation.

\section{Background estimation}
\label{sec:BackgroundEstimation}

\subsection{Background estimation in the four-lepton channel}\label{ss:BackgroundEstimationZZ}

The dominant background to the \HfourL signal comes from $\ZZ/\PZ\PGg^*\to 4\Pell$
production in either $\qqbar$, $\Pg\Pg$, or electroweak processes. 
This background is estimated using the MC simulation discussed in Section~\ref{sec:Simulation}. 
In addition, the \mell selection region between 105 and 140\GeV is wide enough to retain 
sideband regions for further constraints based on the data. 

An additional background to the \Hboson signal, $\PZ+X$ in the following, comes from processes in which decays 
of heavy-flavor hadrons, decays of light mesons within jets, or charged hadrons overlapping with $\pi^0$ 
decays are misidentified as leptons. The main process contributing to these backgrounds is \Zjets, 
while subdominant processes in order of importance are $\ttbar+\text{jets}$, $\PZ\PGg+\text{jets}$, $\PW\PZ+\text{jets}$, and $\PW\PW+\text{jets}$.
The $\PZ+X$ background is estimated from control regions in data. The control regions are defined as events that contain 
a lepton pair satisfying all the requirements of a $\PZ$ candidate and two additional opposite-sign 
leptons where the two additional leptons satisfy identification requirements looser than those used in the analysis. 
These four leptons are then required to pass the \PH candidate selection. The yield in the signal region 
is obtained by weighting the control region events by the lepton misidentification probability, $f_{\Pell}$, defined 
as the fraction of nonsignal leptons that are identified by the analysis selection criteria. A detailed description 
of the method can be found in Ref.~\cite{Sirunyan:2017exp}.

\subsection{Background estimation in the \texorpdfstring{\bbbar channel}{bb}}\label{ss:BackgroundEstimationbb}

The \bbbar channel is used to search for a narrow signal in the \PH boson candidate \mpnet distribution in the T 
and M regions of the $\PGg$-tagged category. The mass distributions in the $\PGg$-tagged category 
are split into two \pt regions (300--400\GeV and $>$400\GeV). Mass distributions of the nonresonant background 
are estimated using a pass-to-fail ratio method, described in the following paragraphs. The other relevant background 
processes in the $\PGg$-tagged category, \Zgamma and \Wgamma are estimated from simulation. 
The simulated \Zgamma distributions are corrected by fitting the \PZ boson candidate \mpnet distributions to 
the data in the Untagged category. The nonresonant background is also estimated with the pass-to-fail ratio method 
in the Untagged category. The \Zjets and \Wjets are estimated from simulation. Simulated distributions of the SM 
$\PH\to\bbbar$  process are also included in the Untagged category, but they have no significant impact on the results 
because of much lower yields compared to other processes.

The pass-to-fail ratio method is based on the ratio of \mpnet distributions between \textsc{ParticleNet} passing ($\text{P}=\text{M}$ or T) 
and failing regions. This gives two pass-to-fail ratios (\rpf) per category. The $R_\text{M/F}(\mpnet,\pt)$, $R_\text{{T/F}}(\mpnet,\pt)$ 
in the $\PGg$-tagged category and $R_\text{{M/F}}^{0\PGg}(\mpnet)$, $R_\text{{T/F}}^{0\PGg}(\mpnet)$ in the Untagged category. 

The nonresonant background, \eg, in the M region, is defined through the relation:
\begin{equation}
n_{\text{M},\text{nonres.}}(i) = n_{F,\text{nonres.}}(i)\,R_\text{M/F}(\mpnet,\pt),
\end{equation}
where $n_{M(F),\text{nonres.}}(i)$ is the number of nonresonant events in the i-th bin in the M (F) region. 
The F region is overwhelmingly dominated by nonresonant events, so $n_{\text{F},\text{nonres.}}(i)$ can be directly estimated 
from data by subtracting the relatively small simulated yields of other processes. The $R_\text{M/F}$ is modeled as 
a polynomial in \mpnet and \pt. In the $\PGg$-tagged category, a polynomial of order $n$ corresponds to 
a linear combination of $\mpnet^a \pt^b$ terms where $a+b\le n$. Furthermore, since there are only two \pt regions, 
a condition $b<2$ is imposed. In the Untagged category, a simple one-dimensional polynomial in \mpnet is used 
to model the corresponding ratio. The polynomial parameters are determined along with other likelihood parameters, 
while simultaneously fitting the model to the data across all regions.
The above also applies to the estimate of nonresonant background in the T region.

A Fisher's F-test~\cite{Fisher} on the unblinded fit to the data is used to determine the minimum polynomial order necessary and sufficient 
for the model. Starting from polynomials of order zero (constant \rpf), terms are
added until no significant improvement is observed. The addition of terms is done independently for all four \rpf. 
The F-test shows that the first-order polynomial is preferred for each of the \rpf.

The prominent \PZ boson peak in the Untagged category is used to measure the \textsc{ParticleNet} efficiency data-to-simulation 
scale factors. The two scale factors (T and M) are free parameters in the joint fit of $\PGg$-tagged and 
Untagged categories. They allow the migration of events from the T and M regions to the F region 
(and vice-versa), controlling the heights of the \PZ boson peak in both event categories. The scale factors are also 
applied to the signal $\PGg \PH$ process.

\section{Systematic uncertainties}
\label{sec:Systematics}

The measurement of the signal strength in this analysis is mainly affected by statistical uncertainties.
Both experimental and theoretical systematic uncertainties that affect the shapes and the yields of the
signals and backgrounds are incorporated in the likelihood as nuisance parameters~\cite{CMS:2024onh}.
The dominant theoretical uncertainties come from the QCD renormalization and factorization scales involved 
in the cross section computation. The uncertainties in the electroweak corrections on the cross sections, 
PDFs, and showering are also included.

Experimental uncertainties common to both the $4\Pell$ and $\bbbar$  channels include uncertainties in the photon identification efficiency,
as well as photon energy scale and resolution.
The normalization of the signal and background processes derived from the MC simulation is affected by the uncertainty 
in the integrated luminosity~\cite{Sirunyan:2759951,CMS-PAS-LUM-17-004,CMS-PAS-LUM-18-002}. 

When the \ZX background is estimated for the four-lepton channel, the flavor composition of QCD-evolved jets 
misidentified as leptons may be different in the $\PZ+1\Pell$ and $\PZ+2\Pell$ control regions, and together with 
the statistical uncertainty in the $\PZ+2\Pell$ region, this uncertainty accounts for about a $\pm$30\% variation 
in the \ZX background~\cite{CMS:2021ugl}.
Uncertainties in the lepton momentum resolution and reconstruction efficiencies are small compared to other uncertainties. 

In the $\bbbar$  channel, the dominant experimental uncertainty comes from the \textsc{ParticleNet} tagging efficiency, yielding 
an uncertainty in the $\PGg \PH$ signal normalization in the signal region of about 15\%~\cite{CMS-PAS-BTV-22-001}. Other notable experimental 
uncertainties are those associated with the jet energy and mass corrections, and the trigger efficiency, 
affecting the signal yield by $<$1\%.

Theoretical systematic uncertainties related to the analysis of light-quark Yukawa couplings are discussed in Section~\ref{sec:HqqResults}, 
along with the theoretical model of the \Hboson production cross section.

\section{\texorpdfstring{Results for \gammaH cross section and \HVV couplings}{Results for gammaH cross section and HVV couplings}}
\label{sec:Results}

The cross section of the \gammaH process, \xsecGammaH, is constrained in the \Htobbbar and $4\Pell$ channels. 
In both cases, a joint fit of the \gammaH-tagged events and Untagged events is performed. 
The distribution of the \mpnet observable in the \bbbar channel is shown in Fig.~\ref{fig:postfit_bb},
and the distribution of the \Dbkg observable in the $4\Pell$ channel is shown in Fig.~\ref{fig:postfit_H4l}. 
The two sets of distributions are shown after a joint fit to the data. No significant excess of \gammaH events is visible in either channel. 

The extended likelihood function is constructed using the probability densities describing the signal and background events
as functions of the \mpnet and \Dbkg observables in the \Htobbbar and $4\Pell$ channels, respectively. 
Simulation or control samples in data estimation are used to describe probability densities, as described in Sec.~\ref{sec:BackgroundEstimation}. 
The \xsecGammaH is parameterized as a function of the couplings \cyy, \czy, \tcyy, and \tczy~\cite{Davis:2021tiv}, 
where the total width \GH of the \Hboson is assumed to have the SM value. 
It has been checked that \GH does not change significantly with variation of $c_i$ within constraints obtained in this analysis,
nor do kinematic distributions in the \Hell decay, which can be assumed to be dominated by the SM tree-level coupling.

\begin{table*}[htb]
\centering
\topcaption{
Observed and expected constraints on the \gammaH cross section \xsecGammaH
and on the \cyy, \czy, \tcyy, and \tczy couplings
using the \Htobbbar and $4\Pell$ channels combined.
The third and fourth rows show constraints on cross section multiplied by the branching fraction
using the \Htobbbar and \Hell channels only, respectively.
The 68\% (central value with uncertainties) and 95\% (upper limit or allowed intervals) \CL intervals are shown.
\label{tab:results}
}
\cmsTable{
\begin{scotch}{llllll}
\vspace{-0.2cm} \\
Parameter   &  Scenario  & \multicolumn{2}{c}{Observed} & \multicolumn{2}{c}{Expected}  \\
\vspace{-0.4cm} \\
 & &  \multicolumn{1}{c}{68\% \CL}  & \multicolumn{1}{c}{95\% \CL}  &  \multicolumn{1}{c}{68\% \CL}  & \multicolumn{1}{c}{95\% \CL}   \\
\vspace{-0.3cm} \\
\hline
\vspace{-0.2cm} \\
$\xsecGammaH\,(\unit{fb})$  & $\czy =\tczy =0$ & $-7.5^{+8.9}_{-8.8}$ & $<$11.8  &  $0.0^{+ 6.3}_{-6.3}$   & $<$16.1   \\
$\xsecGammaH\,(\unit{fb})$ & float all& $-9.3^{+11.3}_{-11.2}$ & $<$16.4  &  $0.0^{+8.0}_{-8.0}$  & $<$21.5   \\
$\xsecGammaH\,\mathcal{B}_{\bbbar}\,(\unit{fb})$ & float all& $-5.4^{+6.5}_{-6.5}$ & $<$9.5  &  $0.0^{+4.6}_{-4.6}$  & $<$12.4   \\
$\xsecGammaH\,\mathcal{B}_{4\Pell}\,(\unit{fb})$ & float all&  ~~~$0.04^{+0.13}_{-0.04}$ & $<$0.32 &  $0.00^{+0.08}_{-0.00}$ &  $<$0.32 \\
\cyy & float all& ~~~$0.00\pm0.50$ &  $[-0.84,0.84]$  &  $0.00\pm0.63$ & $[-0.90,0.90]$   \\
\cyy & fix others &  ~~~$0.00\pm0.46$   &$[-0.77,0.77]$ &  $0.00\pm0.58$  & $[-0.83,0.83]$  \\
\czy & float all&  ~~~$0.00\pm0.18$  & $[-0.30,0.30]$  &  $0.00\pm0.22$  & $[-0.32,0.32]$  \\
\czy & fix others & ~~~$0.00\pm0.16$  & $[-0.27,0.27]$ &  $0.00\pm0.21$  & $[-0.29,0.29]$  \\
\tcyy &  float all& ~~~$0.00\pm0.50$  & $[-0.84,0.84]$ &  $0.00\pm0.63$  & $[-0.90,0.90]$   \\
\tcyy & fix others & ~~~$0.00\pm0.46$  & $[-0.77,0.77]$  &  $0.00\pm0.58$  & $[-0.83,0.83]$    \\
\tczy & float all& ~~~$0.00\pm0.18$  & $[-0.30,0.30]$  &  $0.00\pm0.22$  & $[-0.32,0.32]$    \\
\tczy & fix others & ~~~$0.00\pm0.16$  & $[-0.27,0.27]$  &  $0.00\pm0.21$  & $[-0.29,0.29]$  \\
\vspace{-0.2cm} \\
\end{scotch}
}
\end{table*}

The likelihood ${\cal L}$ is maximized with respect to the nuisance parameters,
describing the systematic uncertainties, and the parameters of interest.
There are four parameters of interest, which can either be the four anomalous couplings 
\cyy, \czy, \tcyy, and \tczy, or the cross-section \xsecGammaH and three of the anomalous couplings, 
with the option to eliminate one of the couplings of choice.
Likelihood maximization is done using the Higgs \textsc{Combine} tool~\cite{CMS:2024onh}.
The allowed 68 and 95\% confidence level (\CL) intervals are defined using the profile likelihood function, 
$-2\Delta{\cal L} =1.00$ and $3.84$, for which exact coverage is expected in the asymptotic limit~\cite{Cowan_2011}. 
The 95\% \CL upper limits on \xsecGammaH are determined with the requirement that cross sections must be positive. 
They are calculated using the \CLs criterion~\cite{Junk:1999kv, 0954-3899-28-10-313} with the modified profiled 
likelihood ratio~\cite{Cowan_2011} as the test statistics.

\begin{figure}[tb!]
    \centering
    \includegraphics[width=0.40\textwidth]{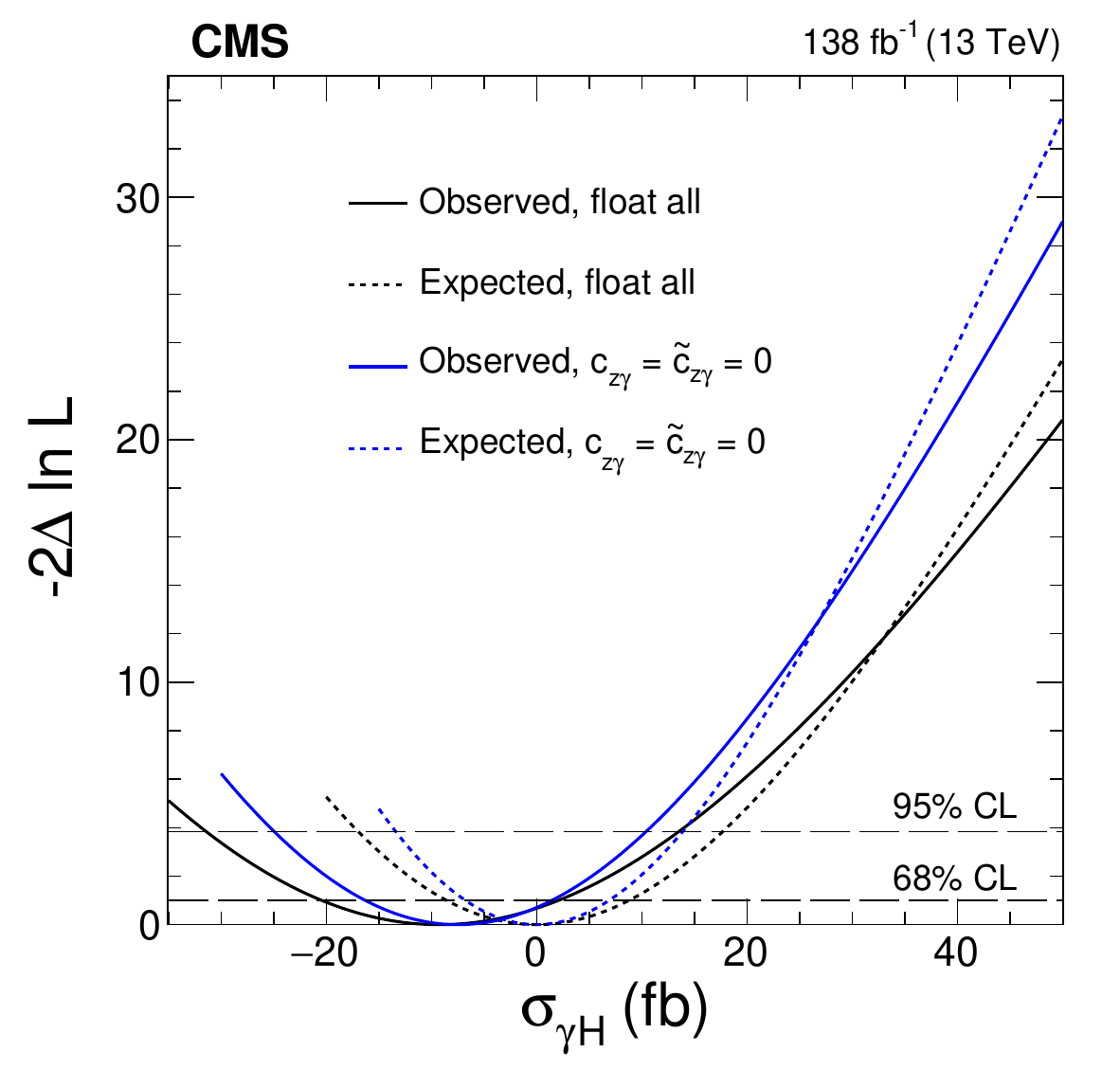}
    \caption{
Constraints on \xsecGammaH from the combination of the \Htobbbar and $4\Pell$ channels.
The results are shown with only \cyy and \tcyy floating in the fit (blue) 
and with all four couplings allowed to float (black).  
Observed (solid) and expected (dashed) likelihood scans are shown. 
The dashed horizontal lines show the 68 and 95\% \CL intervals.
}
    \label{fig:limitsigma}
\end{figure}

Figure~\ref{fig:limitsigma} shows the one-dimensional likelihood scan on \xsecGammaH using the combination of the the \Htobbbar and $4\Pell$ channels.
Due to the much larger branching fraction of the \Htobbbar channel, the combined results are dominated by these decays.
Table~\ref{tab:results} shows a summary of the 68 and 95\% \CL intervals on \xsecGammaH,
either with no constraints on the couplings or allowing only the \cyy and \tcyy couplings.
The results for the \Hell and \Htobbbar channels are presented separately with the cross section multiplied by the respective branching fraction.
This highlights that, when the branching fraction is considered, the \Hell channel is not competitive with the \Htobbbar channel.
Negative cross sections are allowed, as indicated by the central values and 68\% uncertainty ranges, except when setting the 95\% \CL upper limits on \xsecGammaH.
In Fig.~\ref{fig:limitsigma}, the likelihood scan terminates for certain negative values of \xsecGammaH. 
This represents the case where some bins in the observables yield a negative number of events, 
so the total probability density function becomes negative for this extreme case.

\begin{figure}[htb!]
    \centering
    \includegraphics[width=0.40\textwidth]{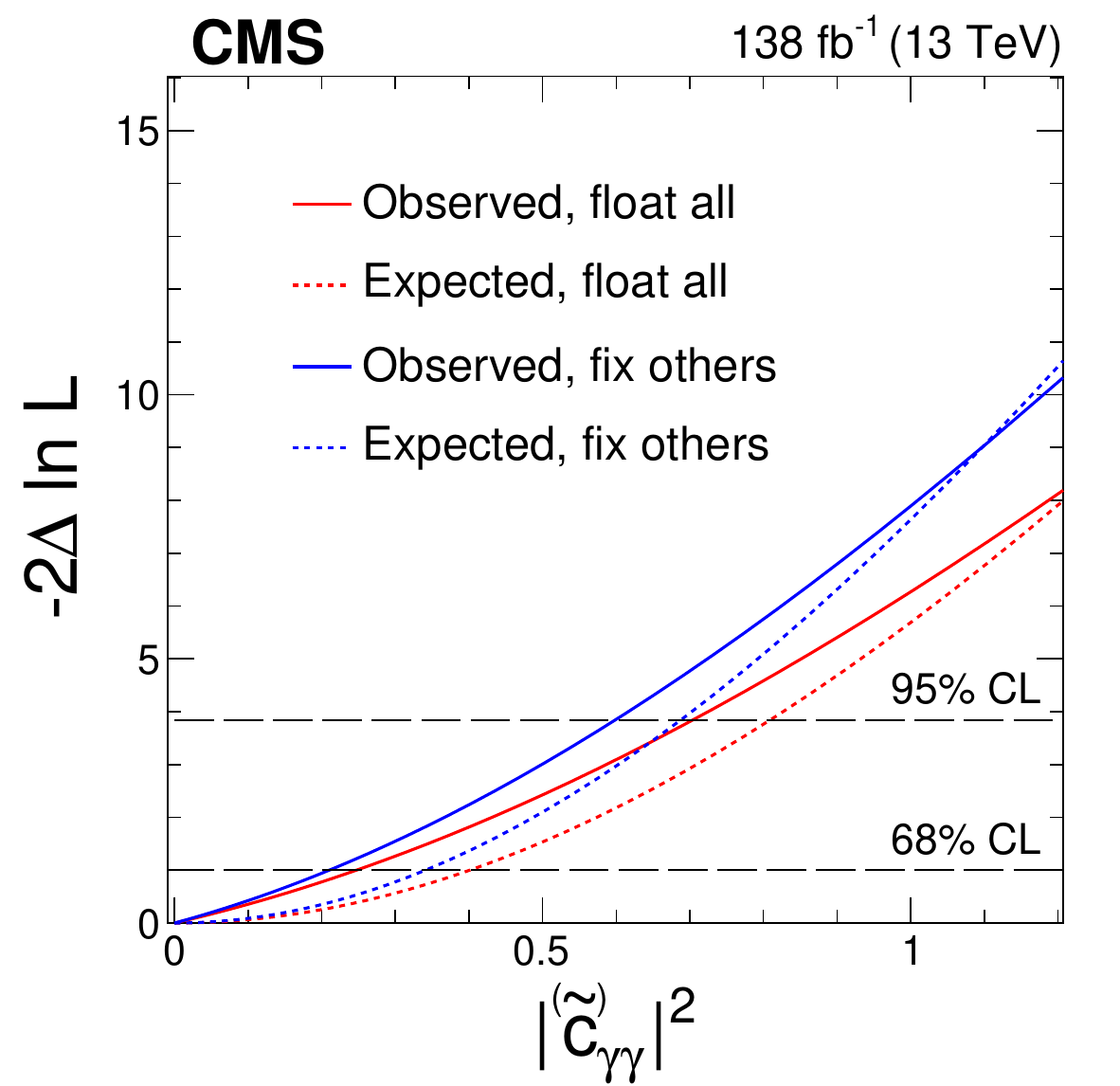}
    \includegraphics[width=0.40\textwidth]{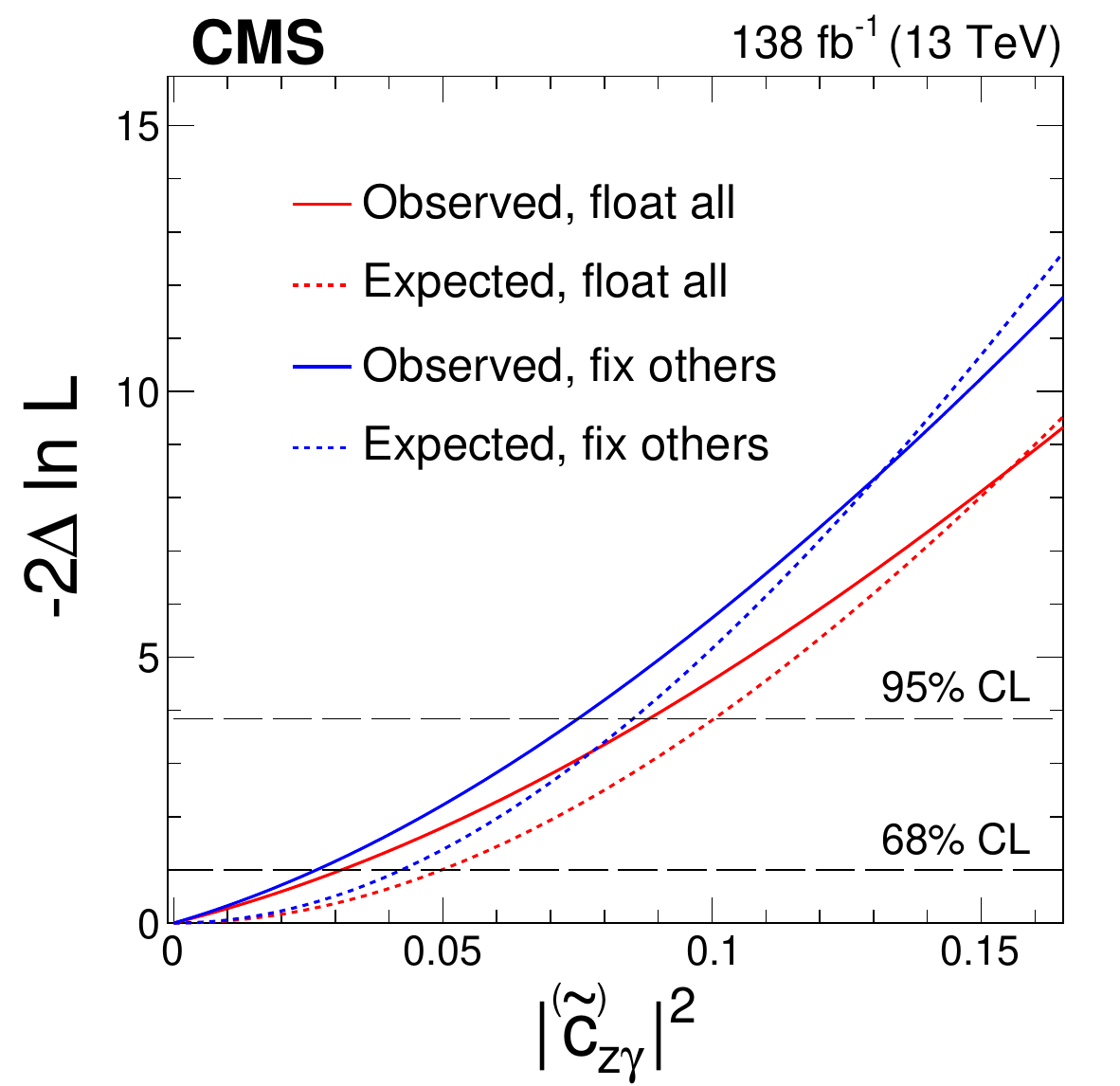}
    \caption{
Constraints on the square of $\abs{\cyy}$ (or $\abs{\tcyy}$) and $\abs{\czy}$ (or $\abs{\tczy}$)
from the combination of the \Htobbbar and $4\Pell$ channels.
The other couplings are either fixed to the null SM expectation (blue) or are left floating in the fit (red).
Observed (solid) and expected (dashed) likelihood scans are shown. 
The dashed horizontal lines show the 68 and 95\% \CL intervals.}
    \label{fig:limitwilsonc}
\end{figure}

Figure~\ref{fig:limitwilsonc} shows the constraints on the squared couplings 
$\cyy^{\,2}$, $\czy^{\,2}$, $\tcyy^{~2}$, and $\tczy^{~2}$.
Due to symmetry between the \CP-even ($\cyy^{\,2}$ or $\czy^{\,2}$) and 
\CP-odd ($\tcyy^{~2}$ or $\tczy^{~2}$) couplings,
the likelihood scans projected on one dimension are indistinguishable, which is why only two graphs are displayed.
We present constraints on the squared values because they are directly related to the event yields and
most closely follow the Gaussian probability distributions. 
Table~\ref{tab:results} shows a summary of the 68 and 95\% \CL intervals on 
four couplings, either with no constraints on the couplings or with certain couplings constrained to the null SM expectation.

\section{Analysis of the light-quark Yukawa couplings}
\label{sec:HqqResults}

The limits on \gammaH production could be used to constrain the Yukawa couplings
of light quarks, as proposed in Ref.~\cite{Aguilar-Saavedra:2020rgo} and discussed in Section~\ref{sec:Introduction}.
However, the photon involved in this process is considerably softer than the one in diagram Fig.~\ref{fig:Feyn1} (left).
This motivates a reoptimization of the photon selection criteria introduced in Section~\ref{ss:EventSelectionZZ}.
Upon testing, we find that the expected constraints remain unchanged, regardless of the selection requirement on $\pt^\PGg$.
This effect is observed because a soft photon can easily be associated with an \Hboson produced via gluon fusion, for instance. 
Therefore, the distinction between the Untagged and $\PGg$-tagged categories is unnecessary.
Instead, we must constrain light-quark Yukawa couplings by examining changes in the \Hboson production rate, 
regardless of the production mode.

This leads us to perform an inclusive analysis of \Hell production, similar to the methods proposed in 
Refs.~\cite{Zhou:2015wra,Zhou:2019}, but with a more detailed focus on both the reconstruction and computational aspects.
The four-lepton final state is especially suitable for this analysis because the decay is largely unaffected by the \Hboson
couplings to quarks at the relevant scale. Additionally, the inclusive \Hell reconstruction maintains high signal purity 
and is almost entirely independent of the production mechanism.
Consequently, only the effects on production couplings and total width \GH of the \Hboson need to be taken into 
account. In this interpretation, 
we combine the Untagged and $\PGg$-tagged categories into one, while keeping the $4\Pell$ selection criteria unchanged.

\subsection{Dependence of the cross sections on the light-quark Yukawa couplings}
\label{sec:scrossdependence}

The SM \bbH process serves as a reference for \Hboson production driven by light quarks at the tree level, \qqbarH,
where the rates are adjusted according to \kappau, \kappad, \kappas, \kappac, and \kappab. 
The rate of the gluon fusion process is also expressed as a function of \kappaq, 
while all other production mechanisms of the \Hboson considered remain unchanged.
The couplings \kappab and \kappat are well constrained by the analysis of the on-shell \Hboson data~\cite{ATLAS:2022vkf,CMS:2022dwd}, 
including the \Htobbbar decay, and \ttH and \ggH production, and thus are fixed to $\kappab = \kappat =1$ in this analysis. 
However, they are sometimes allowed to vary for studies and validation of the techniques presented in Section~\ref{sec:fitYukawa}. 

The total width of the \Hboson is parameterized, as shown below:

\ifthenelse{\boolean{cms@external}}
{
\begin{multline}
    \label{eq:ratio-width}
    \GH = \Rgg(\kappaall)  \, \Gamma^\textrm{SM}_{\Htogg} +\sum_{\PQq=\PQu,\PQd,\PQs,\PQc,\PQb}\kappaq^2 \, \Gamma^\textrm{SM}_{\Htoqqbar}\\
    +\sum_{\PV\PV^\prime}\kappavvprime^2 \, \Gamma^\textrm{SM}_{\HtoVVprime}+\sum_{\Pell}\Gamma^\textrm{SM}_{\PH\to\Pell\Pell} +\GHBSM,
\end{multline}
}
{
\begin{equation}
    \label{eq:ratio-width}
    \GH =
    \Rgg(\kappaall)  \, \Gamma^\textrm{SM}_{\Htogg}
    +\sum_{\PQq=\PQu,\PQd,\PQs,\PQc,\PQb}\kappaq^2 \, \Gamma^\textrm{SM}_{\Htoqqbar}
    +\sum_{\PV\PV^\prime}\kappavvprime^2 \, \Gamma^\textrm{SM}_{\HtoVVprime}
    +\sum_{\Pell}\Gamma^\textrm{SM}_{\PH\to\Pell\Pell}
    +\GHBSM,
    \end{equation}
}

where the partial widths $\Gamma^\textrm{SM}_{\PH\to f}$ are calculated using the SM values for all couplings, 
while the partial width for decays to BSM particles is generally unknown and is constrained only 
by $\GHBSM \ge 0$. 
In the partial width for the decay \HtoVVprime, $\PV$ or $\PV^\prime$ can be $\PW$, $\PZ$, or $\PGg$. 
Since this decay is primarily governed by the tree-level couplings $\PH\PZ\PZ$ and $\PH\PW\PW$, 
the influence of light-quark Yukawa couplings is highly suppressed and 
negligible compared to the direct \Htoqqbar partial width with enhanced couplings.
The coupling strength modifiers \kappavvprime are introduced to account for potential BSM effects 
and will be elaborated on below. In the SM, their values are $\kappavvprime = 1$.
The partial width for the decay $\PH\to\Pell\Pell$, where $\Pell$ denotes leptons $\Pe$, $\mu$, or $\tau$,
and neutrinos are neglected, is independent of \kappaq.
Thus, the dependence on \kappaq is present only in the \Htoqqbar and \Htogg processes.

The cross section of the on-shell \Hboson production at the LHC, $\Pp\Pp\to \HfourL$,
is inversely proportional to the total width of the \Hboson and is parameterized as shown below:

\ifthenelse{\boolean{cms@external}}{
\begin{multline}
    \label{eq:ratio-sigma}
    \sigma_{\Hell} =
    \frac{\Gamma_{\HfourL}^\textrm{SM} \, \kappaZZ^2}{\Gamma_\PH(\kappaall)} \bigg(\Rgg(\kappaall) \, \sigma^\textrm{SM}_{\ggH} \\
    + \sum_{\PQq} \kappaq^2 \, \sigma^\textrm{SM}_{\qqbarH}+ \sigma^\textrm{SM}_{\PQt\PAQt\PH}+ \sigma^\textrm{SM}_{\PQt\PH}+ \sum_{\PV\PV} \kappa_{\PV\PV}^2\, \sigma^\textrm{SM}_{\PV\PV\PH}\bigg),
\end{multline}    
}
{
\begin{equation}
    \label{eq:ratio-sigma}
    \sigma_{\Hell}=
    \frac{\Gamma_{\HfourL}^\textrm{SM} \, \kappaZZ^2}{\Gamma_\PH(\kappaall)}
    \left(
    \Rgg(\kappaall) \, \sigma^\textrm{SM}_{\ggH}
    +\sum_{\PQq}\kappaq^2 \, \sigma^\textrm{SM}_{\qqbarH}
    +\sigma^\textrm{SM}_{\PQt\PAQt\PH}
    +\sigma^\textrm{SM}_{\PQt\PH}
    +\sum_{\PV\PV}\kappa_{\PV\PV}^2 \, \sigma^\textrm{SM}_{\PV\PV\PH}
    \right),
\end{equation}
}

where the partial width for the decay $\PH\to\ZZ\to 4\Pell$, which is a subprocess of $\PH\to\ZZ$,
is scaled by the coupling strength modifier \kappaZZ introduced earlier, 
and all cross sections $\sigma^\mathrm{SM}_i$ are computed for the inclusive on-shell \Hboson production
using the SM values for all couplings.
The production in association with top quarks ($\PQt\PAQt\PH$, $\PQt\PH$) and $\PV\PH$ or 
VBF production ($\PV\PV\PH$) are independent of \kappaq. 
The latter arises for reasons similar to those discussed for the \HtoVVprime decay,
with some exceptions, such as the box diagram in the \ggZH production. 
However, this contribution is minor in the \ggZH process and is negligible compared 
to the $\Pg\Pg\to\PH$ process for any values of the quark Yukawa couplings. 
The VBF, \ZH, and $\PW\PH$ production processes are scaled by either $\kappaZZ^2$ 
or $\kappaWW^2$, with the interference between the two contributions in the VBF process being negligible.

\begin{table*}[htb]
\centering
\topcaption{Central values of the input and derived parameters used in calculations involving Eqs.~(\ref{eq:ratio-width}) and~(\ref{eq:ratio-sigma}).
The list of partons~($p$) comprises gluons ($\Pg$) and five quark flavors ($\PQq$). 
All cross sections $\sigma_i$ are computed for the inclusive on-shell \Hboson production using the SM values for all couplings, 
except for the specific coupling \kappaq that is explicitly mentioned. 
\label{tab:Yukawa-input}
}
\begin{scotch}{llllll}
\vspace{-0.2cm} \\
$p$
    &  $\Gamma^\textrm{SM}_{\PH\to{p\overline{p}}} / \Gamma^\textrm{SM}_{\PH}$  
    & $\sigma^\textrm{SM}_{p\overline{p}\PH}$ (\unit{pb})~~~~~~~
    & $\frac{\sigma_{\ggH}(\kappaq=0)}{\sigma^\textrm{SM}_{\ggH}} - 1$
    & $\frac{\sigma_{\ggH}(\kappaq\gg1)}{\sigma^\textrm{SM}_{\ggH}}$ 
    & $\frac{\sigma_{\qqbarH}(\kappaq\gg1)}{\sigma^\textrm{SM}_{\ggH}}$  \\
\vspace{-0.2cm} \\
\hline
\vspace{-0.2cm} \\
$\Pg$ & $8.187 \times 10^{-2}$ & $4.858 \times 10^{}$ & \NA & \NA & \NA \\
$\PQb$ & $5.824 \times 10^{-1}$ & $4.880 \times 10^{-1}$ & $1.595$ & $1.422 \times 10^{-2}$ & $1.723 \times 10^{-2}$ \\
$\PQc$ & $2.891 \times 10^{-2}$ & $7.735 \times 10^{-2}$ & $4.254 \times 10^{-2}$ & $2.794 \times 10^{-3}$ & $5.506 \times 10^{-2}$ \\
$\PQs$ & $2.152 \times 10^{-4}$ & $1.854 \times 10^{-3}$ & $5.040 \times 10^{-4}$ & $1.518 \times 10^{-4}$ & $1.774 \times 10^{-1}$ \\
$\PQd$ & $5.552 \times 10^{-7}$ & $1.381 \times 10^{-5}$ & $2.087 \times 10^{-6}$ & $1.459 \times 10^{-6}$ & $5.120 \times 10^{-1}$ \\
$\PQu$ & $1.183 \times 10^{-7}$ & $4.155 \times 10^{-6}$ & $5.050 \times 10^{-7}$ & $4.189 \times 10^{-7}$ & $7.234 \times 10^{-1}$ \\
\vspace{-0.2cm} \\
\end{scotch}
\end{table*}

The rates of the gluon fusion process in Eq.~(\ref{eq:ratio-sigma}) and of the decay to gluons in Eq.~(\ref{eq:ratio-width}) 
are both scaled by the same factor \Rgg, which depends on the couplings of all quarks involved
in the \Hgg loop. 
The calculation of $\Rgg(\kappaall)$ 
is carried out using \MCFM code within 
the \jhugen package, extending the computation in Ref.~\cite{Davis:2021tiv} to include light quarks.
Since \Rgg is used as a scaling factor in front of $\sigma^\textrm{SM}_{\ggH}$ calculated at next-to-NNLO in QCD~\cite{deFlorian:2016spz},
the \ggH cross section in Eq.~(\ref{eq:ratio-sigma}) corresponds to the same order 
under the assumption that the $K$ factor for matching the LO \ggH cross section is independent of the quark flavor.

The partial decay width $\Gamma^\textrm{SM}_{\Htoqqbar}$ in Eq.~(\ref{eq:ratio-width})
is calculated using Ref.~\cite{Spira:1997dg} for $\PQq=\PQu,\PQd,\PQs,\PQc$,
while the \Htobbbar value is obtained from Ref.~\cite{deFlorian:2016spz}.
The decay width of the \Hboson to two quarks is well understood under the assumption that the 
quark masses are small compared to the \Hboson mass $\mH = 125\GeV$, 
which is a valid approximation for the light quarks considered in this analysis. 
Calculations use the running mass of the light quarks at $\mH$~\cite{Huang:2020hdv}. 
A uniform $K$ factor is applied to match the $\PH\to\ccbar$ partial decay width to the most accurate 
value in Ref.~\cite{deFlorian:2016spz}.

The cross section of the \qqbarH process as a function of the light-quark Yukawa coupling
for each flavor of $\PQq$ in Eq.~(\ref{eq:ratio-sigma}) is calculated at NNLO in QCD with SusHi~\cite{Harlander:2012pb}.
As such, this process involves light quarks coupled to the \Hboson in either the initial or final states.
Reference~\cite{Harlander:2015xur} demonstrated that the calculation for \bbH at NNLO in QCD 
is analogous to the calculation for \qqbarH production with light quarks. The main differences stem 
from the distinct quark Yukawa couplings and variations in the flavor composition of the PDFs.
The central values for the QCD factorization and renormalization scales are set to $\mH/4$ and $\mH$, respectively, 
with systematic variations by a factor of 2, leading to cross section uncertainties between 3 and 5\%, 
depending on the quark flavor. Similarly, PDF variations lead to uncertainties between 3 and 9\%.
For the \bbH and $\ccbar\PH$ processes, the cross sections from Ref.~\cite{deFlorian:2016spz} 
are reproduced with good accuracy. 
The typical $K$ factor relative to the LO in the QCD calculation of the process $\PQq\PAQq\to\PH$~is~1.2.
The effect of the interference between the \qqbarH process and gluon fusion at higher orders in QCD 
is found to be negligible compared to the individual contributions.

Table~\ref{tab:Yukawa-input} illustrates the numerical values of the parameters used in calculations 
involving Eqs.~(\ref{eq:ratio-width}) and~(\ref{eq:ratio-sigma}), as well as their asymptotic behavior.
The concept of this analysis can be grasped by examining how the calculation in Eq.~(\ref{eq:ratio-sigma})
varies with \kappaq for a specific quark flavor ${\PQq}$.
Cross sections for all processes that do not explicitly depend on \kappaq will decrease as 
\kappaq increases above 1, due to the rise in $\Gamma_\PH(\kappaq)$ in the denominator, 
as indicated by Eq.~(\ref{eq:ratio-width}). 
The cross section for the gluon fusion process will also decrease, but at a slower rate due to the interplay 
between the quadratic dependence on \kappaq in both the total width and \Rgg.
This cross section will ultimately reach a plateau at $\kappaq\gg1$ when the $\kappaq^2$ 
term becomes dominant in $\Rgg(\kappaq)$.
In contrast, the cross section for the \qqbarH process will increase with rising \kappaq 
because the numerator in Eq.~(\ref{eq:ratio-sigma}) grows faster than the denominator. 
However, this increase will eventually level off, reaching an asymptotic value at $\kappaq\gg1$, 
where the terms proportional to $\kappaq^2$ become dominant in both the numerator and the denominator.

\subsection{Results for light-quark Yukawa couplings}
\label{sec:fitYukawa}

This analysis is feasible because the total cross section for the $\Pp\Pp\to \HfourL$ process decreases 
monotonically with increasing $\kappaq^2$, eventually becoming inconsistent with the data and thus 
placing constraints on \kappaq. Consequently, it is possible to simultaneously set constraints on the 
couplings of all four quark flavors $\PQq = \PQu, \PQd, \PQs, \PQc$, without compromising precision,
provided that the constraints are consistent with null measurements of these couplings. 
This occurs because, in the case of a null result for one coupling, introducing a non-zero value for another 
coupling can only reduce the expected cross section, thereby tightening the constraint.
Once the precision of these measurements is sufficient to detect significant nonzero values, 
distinguishing between the couplings will become challenging.

Incorporating BSM contributions, with \GHBSM in Eq.~(\ref{eq:ratio-width})
and $\kappa_{ZZ}$ in Eq.~(\ref{eq:ratio-sigma}), does not affect the upper limits, given the following two reasons. 
First, introducing a nonnegative \GHBSM will further decrease the expected cross section, 
making its null value the most probable outcome when $\kappaq^2$ is enhanced. 
Second, enforcing the assumption $\kappaZZ^2\le1$ and $\kappaWW^2\le1$ in Eq.~(\ref{eq:ratio-sigma}) 
is theoretically well-supported across various models~\cite{LHCHiggsCrossSectionWorkingGroup:2013rie}.
This includes models with any number of Higgs doublets, both with and without additional Higgs singlets, 
as well as certain types of composite Higgs models. 

Since all four \kappavvprime parameters must be taken into account in Eq.~(\ref{eq:ratio-width}), 
we set $\kappa_{\PZ\PGg}=\kappa_{\PGg\PGg}=1$, $\kappaZZ=\kappaWW$, and $\kappaZZ^2\le1$. 
With this assumption, it is apparent that allowing \kappaZZ to be a free parameter in Eq.~(\ref{eq:ratio-sigma}) 
would result in $\kappaZZ^2 = 1$ being the most likely value when $\kappaq^2$ is increased.
It should be noted that technically the cross section $\sigma^\textrm{SM}_{\PQt\PH}$ in Eq.~(\ref{eq:ratio-sigma}) 
also depends on $\kappaWW$. However, this contribution is insufficient to counteract the reduction 
in other terms in Eq.~(\ref{eq:ratio-sigma}) that results from a decrease in $\kappaZZ^2$.

\begin{figure*}[htb]
\centering
\begin{tabular}{cc}
\includegraphics[width=0.40\textwidth]{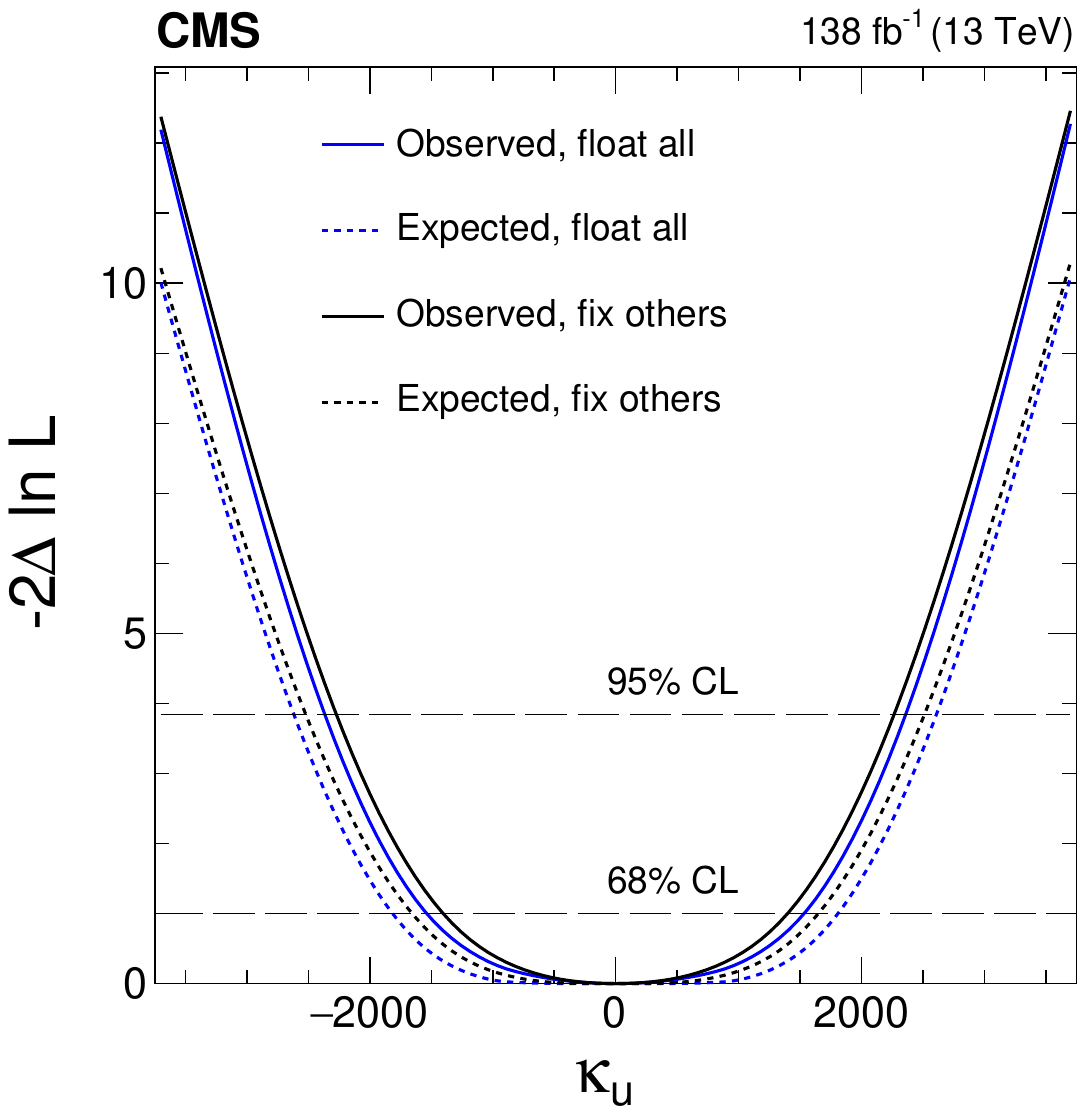}
&
\includegraphics[width=0.40\textwidth]{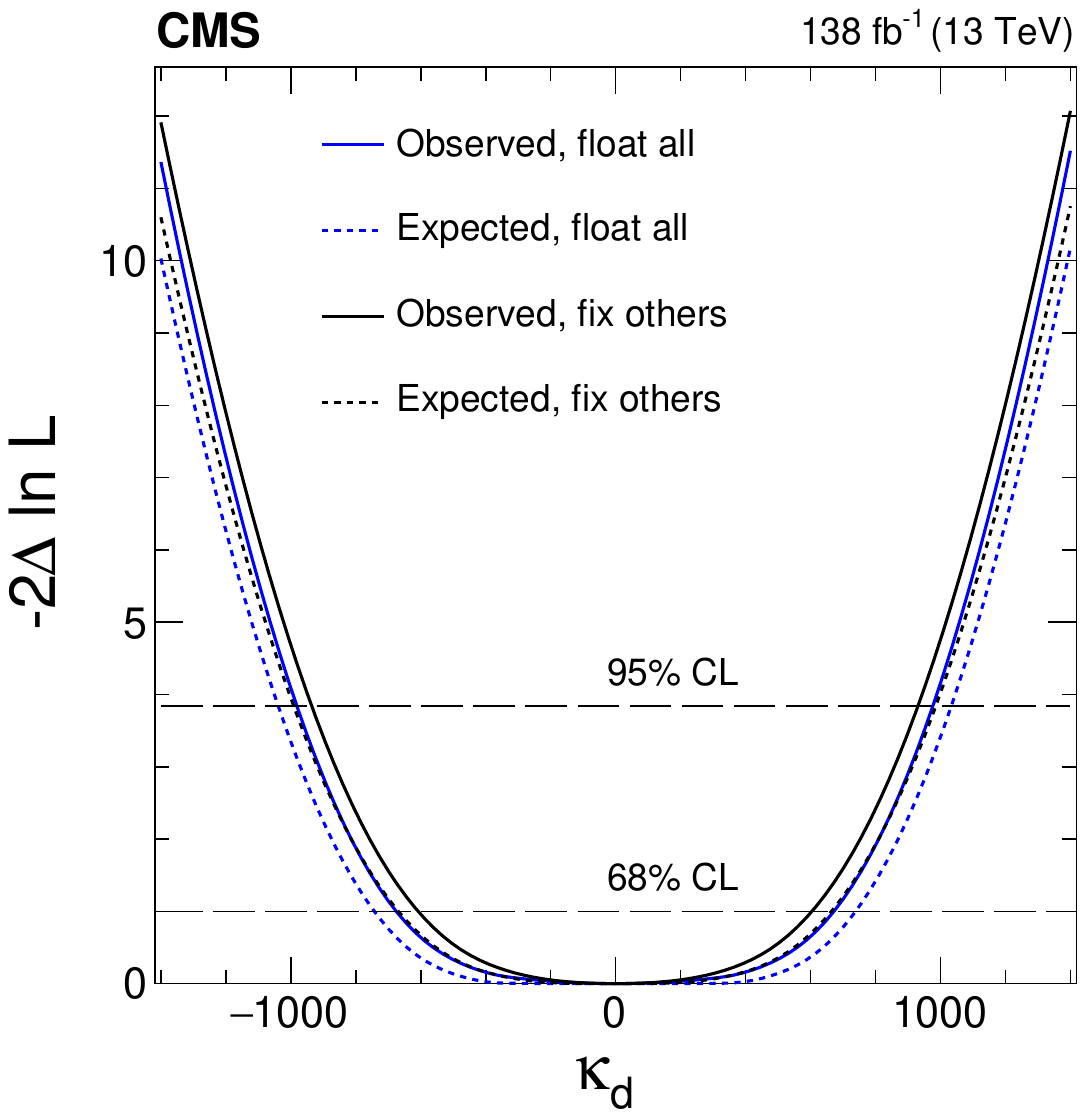} 
\\
\includegraphics[width=0.40\textwidth]{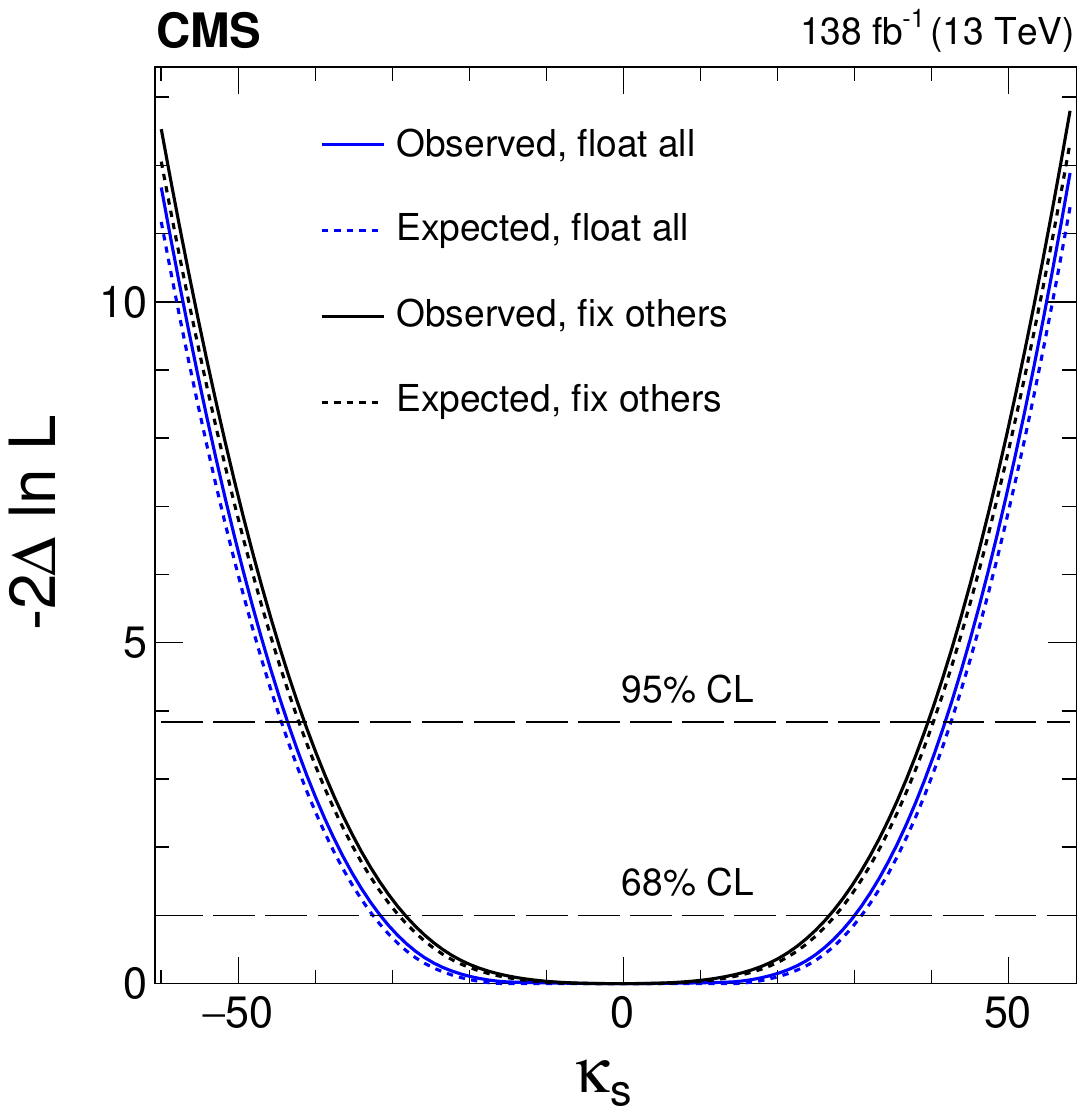}
&
\includegraphics[width=0.40\textwidth]{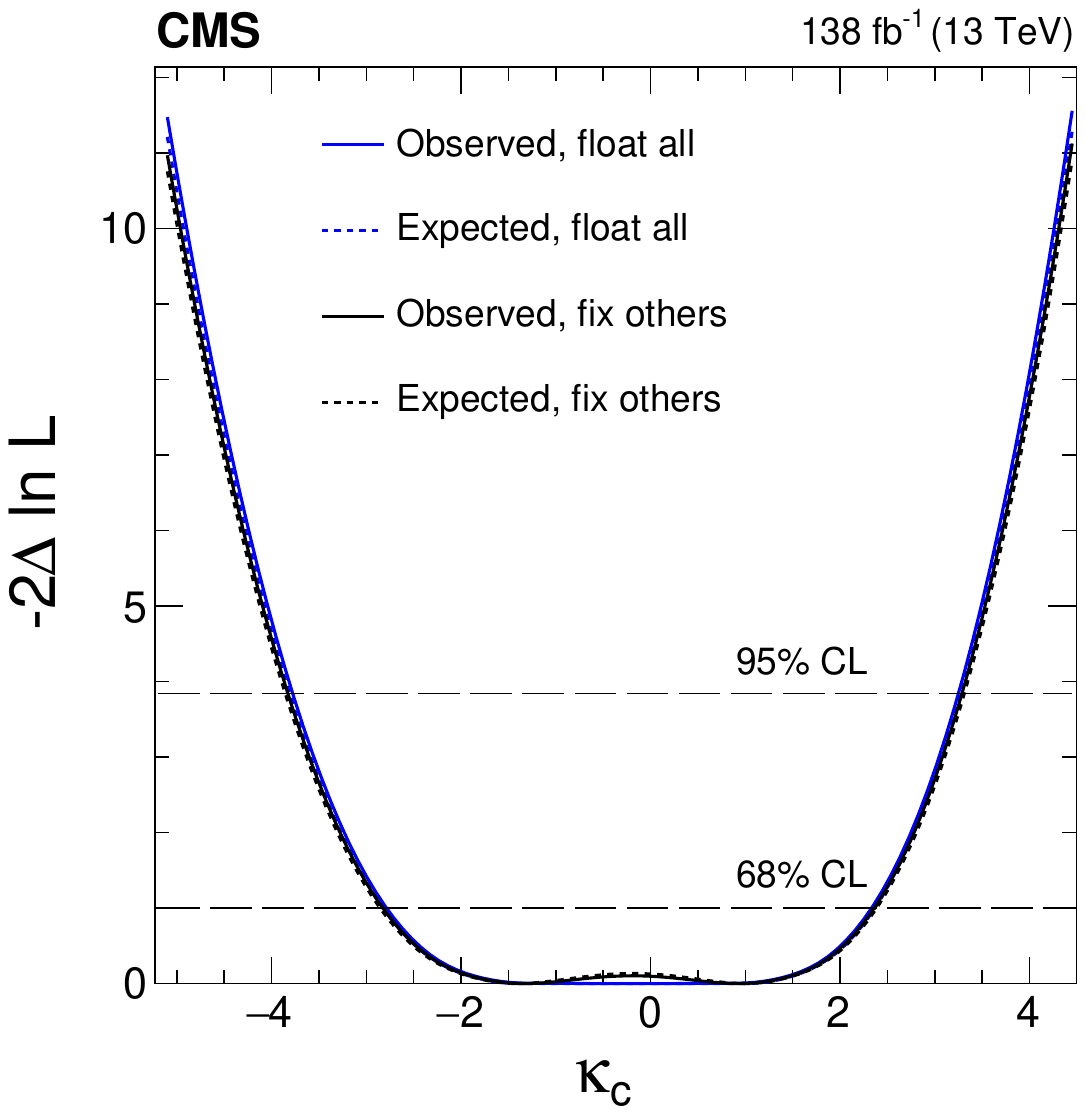}
\\
\end{tabular}
\caption
{
Constraints on \kappau, \kappad, \kappas, and \kappac are shown using the \Hell channel. 
In scenario one (black), all couplings except the one being shown are fixed at their SM values. 
In scenario two (blue), the Yukawa couplings for the three other light quarks are left unconstrained, 
and BSM contributions are allowed: $\kappaZZ^2\le1$ and $\GHBSM\ge0$.
Both observed (solid) and expected (dashed) constraints are presented. 
The crossings of dashed horizontal lines and the likelihood curves indicate the 68 and 95\% \CL intervals.
}
\label{fig:Yukawa-results}
\end{figure*}

\begin{table*}[htb]
\centering
\topcaption{
Observed and expected constraints on the \kappau, \kappad, \kappas, and \kappac couplings
are shown using the \Hell channel. 
In one scenario, all couplings except the one being shown are fixed at their SM values. 
In the other scenario, the Yukawa couplings for the three other light quarks are left unconstrained, and BSM contributions are allowed. 
The 68\% (central value with error bars) and 95\% (bracketed range or upper limit) \CL intervals are displayed.
\label{tab:Yukawa-results}
}
\cmsTable{
\begin{scotch}{llllll}
\vspace{-0.2cm} \\
    Parameter   &  Scenario                & \multicolumn{2}{c}{Observed} & \multicolumn{2}{c}{Expected}  \\
\vspace{-0.4cm} \\
 & &  \multicolumn{1}{c}{68\% \CL}  & \multicolumn{1}{c}{95\% \CL}  &  \multicolumn{1}{c}{68\% \CL}  & \multicolumn{1}{c}{95\% \CL}   \\
\vspace{-0.3cm} \\
\hline
\vspace{-0.2cm} \\
\kappau & float all & $(0.0\pm1.5)\times10^{3}$ & $[-2.4, 2.4]\times10^{3}$ & $(0.0\pm1.8)\times10^{3}$ & $[-2.6, 2.6]\times10^{3}$ \\
\kappau & fix others         & $(0.0\pm1.4)\times10^{3}$ & $[-2.3, 2.3]\times10^{3}$ & $(0.0\pm1.6)\times10^{3}$ & $[-2.5, 2.5]\times10^{3}$  \\

\kappad & float all & $(0.0^{+6.7}_{-6.8})\times10^{2}$ & $[-9.8, 9.8]\times10^{2}$ & $(0.0\pm7.4)\times10^{2}$ & $[-1.0, 1.0]\times10^{3}$ \\
\kappad & fix others         & $(0.0^{+6.0}_{-6.0})\times10^{2}$ & $[-9.4, 9.3]\times10^{2}$ & $(0.0^{+6.6}_{-6.7})\times10^{2}$ & $[-9.9, 9.9]\times10^{2}$  \\

\kappas & float all & $0^{+30}_{-32}$ & $[-44 , 42]$ & $1^{+31}_{-32}$ & $[-44 , 42]$ \\
\kappas & fix others         & $0^{+27}_{-27}$ & $[-41 , 40]$ &  $1^{+27}_{-30}$ & $[-42 , 40]$  \\

\kappac & float all & $0.0^{+2.3}_{-2.8}$ & $[-3.7 , 3.2]$ &  $1.0^{+1.4}_{-3.8}$ & $[-3.8, 3.3]$ \\
\kappac & fix others         & $0.9^{+1.4}_{-3.8}$ & $[-3.8 , 3.2]$ & $1.0^{+1.4}_{-3.8}$ & $[-3.9, 3.3]$ \\

$\GHBSM\,(\MeV)$  &  float all & $0.0^{+0.7}_{-0.0}$ & $<$1.4 & $0.0^{+0.7}_{-0.0}$ & $<$1.4 \\
\vspace{-0.2cm} \\
\end{scotch}
}
\end{table*}

\begin{table*}[htb]
\centering
\topcaption{
Observed and expected constraints on the $\barkappau$, $\barkappad$, $\barkappas$, and $\barkappac$ 
defined as $\barkappaq=y_{\PQq}v/m_{\PQb}$, following the same conventions as outlined in Table~\ref{tab:Yukawa-results}. 
\label{tab:Yukawa-bar}
}
\begin{scotch}{llllll}
\vspace{-0.2cm} \\
    Parameter   &  Scenario                & \multicolumn{2}{c}{Observed} & \multicolumn{2}{c}{Expected}  \\
\vspace{-0.4cm} \\
 & &  \multicolumn{1}{c}{68\% \CL}  & \multicolumn{1}{c}{95\% \CL}  &  \multicolumn{1}{c}{68\% \CL}  & \multicolumn{1}{c}{95\% \CL}   \\
\vspace{-0.3cm} \\
\hline
\vspace{-0.2cm} \\
$\barkappau$ & float all & $0.00\pm0.66$ & $[-1.02, 1.02]$ & $0.00\pm0.79$ & $[-1.13, 1.13]$ \\
$\barkappau$ & fix others         & $0.00^{+0.60}_{-0.61}$ & $[-0.98, 0.98]$ & $0.00^{+0.71}_{-0.72}$ & $[-1.09, 1.09]$  \\

$\barkappad$ & float all & $0.00^{+0.63}_{-0.64}$ & $[-0.93, 0.92]$ & $0.00\pm0.70$ & $[-0.98, 0.97]$ \\
$\barkappad$ & fix others         & $0.00\pm{0.57}$ & $[-0.88, 0.88]$ & $0.00^{+0.63}_{-0.63}$ & $[-0.93, 0.93]$  \\

$\barkappas$ & float all & $0.00^{+0.58}_{-0.60}$ & $[-0.89 , 0.85]$ & $0.02^{+0.57}_{-0.64}$ & $[-0.85 , 0.81]$ \\
$\barkappas$ & fix others         & $-0.01^{+0.52}_{-0.53}$ & $[-0.79 , 0.75]$ &  $0.02^{+0.51}_{-0.58}$ & $[-0.81 , 0.77]$  \\

$\barkappac$ & float all & $0.02^{+0.49}_{-0.63}$ & $[-0.82 , 0.71]$ &  $0.22^{+0.30}_{-0.84}$ & $[-0.83, 0.72]$ \\
$\barkappac$ & fix others         & $0.2^{+0.31}_{-0.83}$ & $[-0.84 , 0.72]$ & $0.22^{+0.30}_{-0.85}$ & $[-0.85, 0.73]$ \\
\vspace{-0.2cm} \\
\end{scotch}
\end{table*}

The fit to the data closely follows the approach outlined in Section~\ref{sec:Results}.
The extended likelihood function is built using probability densities that characterize both signal and background events 
as functions of the \Dbkg observable in the $4\Pell$ channel.
The \Dbkg parameterizations for all \qqbarH processes are modeled based on the ${\PQb\PAQb\PH}$ simulation, 
with the yields of all processes rescaled according to Eq.~(\ref{eq:ratio-sigma}). 
The fit is conducted in two scenarios. In the first scenario, \kappaq for a specific quark flavor ${\PQq}$ is left unconstrained, 
while all other coupling modifiers are fixed at their SM values: $\kappa_{\PQq^\prime}=1$ for $\PQq^\prime\ne\PQq$, 
$\kappaZZ=1$, and $\GHBSM=0$. 
In the second scenario, $\kappab=\kappat=1$ is set, while the Yukawa couplings of all other quarks remain 
unconstrained, with $\kappaZZ^2\le1$ and $\GHBSM\ge0$. 
Figure~\ref{fig:Yukawa-results} illustrates the constraints on each \kappaq, and 
Table~\ref{tab:Yukawa-results} presents the 68 and 95\% \CL intervals for \kappaq in both scenarios.
The results are quite similar in both scenarios, as previously discussed, 
but the results from the simultaneous fitting of all light-quark Yukawa couplings are more general.

The results of the fit, represented by the $\barkappaq$ parameters,
provide a means to compare the hierarchy of Yukawa couplings of light quarks relative to the ${\PQb}$ and ${\PQt}$ quarks.
These results are presented in Table~\ref{tab:Yukawa-bar}.
They are based on the assumption that both third-generation quarks, ${\PQb}$ and ${\PQt}$, 
couple to the \Hboson with strengths consistent with the SM.
Under this assumption, the hypothesis that 
$y_{\PQu}=y_{\PQt}^\mathrm{\,SM}$, $y_{\PQc}=y_{\PQt}^\mathrm{\,SM}$, 
$y_{\PQd}=y_{\PQb}^\mathrm{\,SM}$, or $y_{\PQs}=y_{\PQb}^\mathrm{\,SM}$, 
that is up-type (${\PQu}$ or ${\PQc}$) or down-type (${\PQd}$ or ${\PQs}$) quarks in the first or second generation,
have the same couplings as those in the third generation (${\PQt}$ or ${\PQb}$, respectively), 
is excluded with a \CL greater than 95\%. 
It is not surprising that the limits on $\barkappaq$ for the four light quarks are of a similar magnitude to the SM 
value for the $\PQb$ quark, as it is the Yukawa couplings that make a significant contribution to the \Hboson decay width.

When \GHBSM is allowed to vary in the fit, the resulting constraints are: 
$\GHBSM=0.0^{+0.7}_{-0.0}\MeV$ with an upper limit of $1.4\MeV$ at 95\% \CL.
The constraints are expected to be $\GHBSM=0.0^{+0.7}_{-0.0}\MeV$ $(<$\,$1.4\MeV)$.
This constraint is possible due to the assumptions made about other couplings, such as $\kappaZZ^2 < 1$. 
Bounds on \GHBSM can be obtained from existing off-shell \Hboson data~\cite{CMS:2022ley,ATLAS:2023dnm,CMS:2024offshell} 
without needing constraints on other couplings. However, these bounds, along with those obtained 
from the combined analysis of \Hboson data at the LHC~\cite{ATLAS:2022vkf,CMS:2022dwd}
are valid only under the assumption of small Yukawa couplings for the light quarks. 
This assumption is not applicable to the results presented in this paper.

In all of the fits mentioned above, the third-generation quark couplings are held fixed at their SM values, 
$\kappat=1$ and $\kappab=1$. If these constraints are relaxed to Gaussian constraints with 
11 and 17\% uncertainties, respectively, representing recent constraints from combined CMS data~\cite{CMS:2022dwd}, 
the achieved bounds at 68\% \CL on \kappac, \kappas, \kappad, 
\kappau, and  \GHBSM vary by no more than 47, 37, 33, 33, and 95\%, respectively.
Alternatively, one can constrain only the top quark coupling, such as setting $\kappat=1$, 
while allowing \kappab to vary freely, and still obtain bounds on the light-quark Yukawa couplings 
that are only somewhat less stringent than those shown in Table~\ref{tab:Yukawa-results}.
For instance, a fit where \kappab and \kappac are allowed to vary would be analogous 
to the results presented in Eq.~(12) and Fig.~27~(left) of Ref.~\cite{CMS:2023gjz} using the same \Hell data, 
where only minor approximations were made compared to the methodology used in this paper.

\section{Summary}
\label{sec:Summary}

A search for \gammaH production is performed with the data from the CMS experiment at the LHC 
corresponding to an integrated luminosity of 138\fbinv at a proton-proton center-of-mass collision energy of 13\TeV.
The analysis focuses on the topology of a boosted \Hboson recoiling against a high-energy photon. 
The final states of $\PH\to \bbbar$ and \Hell are analyzed. 
This study examines effective \HZy and \Hyy anomalous couplings within the context of an effective field theory.
In this approach, the observed (expected) constraint on the \gammaH production cross section is
$\xsecGammaH<16.4\,(21.5)\unit{fb}$ at 95\% \CL.
Simultaneous constraints on four anomalous couplings involving \HZy and \Hyy are provided.

Additionally, the production rate for \Hell is examined to assess potential enhancements in the Yukawa couplings 
between light quarks and the \Hboson. 
This includes examining modifications to both direct quark-antiquark annihilation and gluon fusion loop processes.
Assuming the standard model Yukawa couplings for the bottom and top quarks ($\kappab=\kappat=1$),
along with the constraints on the \HVV couplings ($\kappaWW^2\le1$ and $\kappaZZ^2\le1$),
the following simultaneous constraints are obtained:  
$\kappa_\PQu=(0.0\pm1.5)\times10^{3}$, $\kappa_\PQd=(0.0^{+6.7}_{-6.8})\times10^{2}$, $\kappa_\PQs=0^{+30}_{-32}$, and $\kappa_\PQc=0.0^{+2.3}_{-2.8}$.
The hypothesis that up- or down-type quarks in the first or second generation have the same Yukawa couplings 
as those in the third generation is excluded with a \CL greater than 95\%.

\begin{acknowledgments}
    We congratulate our colleagues in the CERN accelerator departments for the excellent performance of the LHC and thank the technical and administrative staffs at CERN and at other CMS institutes for their contributions to the success of the CMS effort. In addition, we gratefully acknowledge the computing centers and personnel of the Worldwide LHC Computing Grid and other centers for delivering so effectively the computing infrastructure essential to our analyses. Finally, we acknowledge the enduring support for the construction and operation of the LHC, the CMS detector, and the supporting computing infrastructure provided by the following funding agencies: SC (Armenia), BMBWF and FWF (Austria); FNRS and FWO (Belgium); CNPq, CAPES, FAPERJ, FAPERGS, and FAPESP (Brazil); MES and BNSF (Bulgaria); CERN; CAS, MoST, and NSFC (China); MINCIENCIAS (Colombia); MSES and CSF (Croatia); RIF (Cyprus); SENESCYT (Ecuador); ERC PRG, RVTT3 and MoER TK202 (Estonia); Academy of Finland, MEC, and HIP (Finland); CEA and CNRS/IN2P3 (France); SRNSF (Georgia); BMBF, DFG, and HGF (Germany); GSRI (Greece); NKFIH (Hungary); DAE and DST (India); IPM (Iran); SFI (Ireland); INFN (Italy); MSIP and NRF (Republic of Korea); MES (Latvia); LMTLT (Lithuania); MOE and UM (Malaysia); BUAP, CINVESTAV, CONACYT, LNS, SEP, and UASLP-FAI (Mexico); MOS (Montenegro); MBIE (New Zealand); PAEC (Pakistan); MES and NSC (Poland); FCT (Portugal);  MESTD (Serbia); MICIU/AEI and PCTI (Spain); MOSTR (Sri Lanka); Swiss Funding Agencies (Switzerland); MST (Taipei); MHESI and NSTDA (Thailand); TUBITAK and TENMAK (Turkey); NASU (Ukraine); STFC (United Kingdom); DOE and NSF (USA).
       
    \hyphenation{Rachada-pisek} Individuals have received support from the Marie-Curie program and the European Research Council and Horizon 2020 Grant, contract Nos.\ 675440, 724704, 752730, 758316, 765710, 824093, 101115353, 101002207, and COST Action CA16108 (European Union); the Leventis Foundation; the Alfred P.\ Sloan Foundation; the Alexander von Humboldt Foundation; the Science Committee, project no. 22rl-037 (Armenia); the Fonds pour la Formation \`a la Recherche dans l'Industrie et dans l'Agriculture (FRIA-Belgium); the Beijing Municipal Science \& Technology Commission, No. Z191100007219010 and Fundamental Research Funds for the Central Universities (China); the Ministry of Education, Youth and Sports (MEYS) of the Czech Republic; the Shota Rustaveli National Science Foundation, grant FR-22-985 (Georgia); the Deutsche Forschungsgemeinschaft (DFG), among others, under Germany's Excellence Strategy -- EXC 2121 ``Quantum Universe" -- 390833306, and under project number 400140256 - GRK2497; the Hellenic Foundation for Research and Innovation (HFRI), Project Number 2288 (Greece); the Hungarian Academy of Sciences, the New National Excellence Program - \'UNKP, the NKFIH research grants K 131991, K 133046, K 138136, K 143460, K 143477, K 146913, K 146914, K 147048, 2020-2.2.1-ED-2021-00181, TKP2021-NKTA-64, and 2021-4.1.2-NEMZ\_KI-2024-00036 (Hungary); the Council of Science and Industrial Research, India; ICSC -- National Research Center for High Performance Computing, Big Data and Quantum Computing and FAIR -- Future Artificial Intelligence Research, funded by the NextGenerationEU program (Italy); the Latvian Council of Science; the Ministry of Education and Science, project no. 2022/WK/14, and the National Science Center, contracts Opus 2021/41/B/ST2/01369 and 2021/43/B/ST2/01552 (Poland); the Funda\c{c}\~ao para a Ci\^encia e a Tecnologia, grant CEECIND/01334/2018 (Portugal); the National Priorities Research Program by Qatar National Research Fund; MICIU/AEI/10.13039/501100011033, ERDF/EU, "European Union NextGenerationEU/PRTR", and Programa Severo Ochoa del Principado de Asturias (Spain); the Chulalongkorn Academic into Its 2nd Century Project Advancement Project, and the National Science, Research and Innovation Fund via the Program Management Unit for Human Resources \& Institutional Development, Research and Innovation, grant B39G670016 (Thailand); the Kavli Foundation; the Nvidia Corporation; the SuperMicro Corporation; the Welch Foundation, contract C-1845; and the Weston Havens Foundation (USA).
\end{acknowledgments}

\bibliography{auto_generated}
\cleardoublepage \appendix\section{The CMS Collaboration \label{app:collab}}\begin{sloppypar}\hyphenpenalty=5000\widowpenalty=500\clubpenalty=5000\cmsinstitute{Yerevan Physics Institute, Yerevan, Armenia}
{\tolerance=6000
V.~Chekhovsky, A.~Hayrapetyan, V.~Makarenko\cmsorcid{0000-0002-8406-8605}, A.~Tumasyan\cmsAuthorMark{1}\cmsorcid{0009-0000-0684-6742}
\par}
\cmsinstitute{Institut f\"{u}r Hochenergiephysik, Vienna, Austria}
{\tolerance=6000
W.~Adam\cmsorcid{0000-0001-9099-4341}, J.W.~Andrejkovic, L.~Benato\cmsorcid{0000-0001-5135-7489}, T.~Bergauer\cmsorcid{0000-0002-5786-0293}, S.~Chatterjee\cmsorcid{0000-0003-2660-0349}, K.~Damanakis\cmsorcid{0000-0001-5389-2872}, M.~Dragicevic\cmsorcid{0000-0003-1967-6783}, P.S.~Hussain\cmsorcid{0000-0002-4825-5278}, M.~Jeitler\cmsAuthorMark{2}\cmsorcid{0000-0002-5141-9560}, N.~Krammer\cmsorcid{0000-0002-0548-0985}, A.~Li\cmsorcid{0000-0002-4547-116X}, D.~Liko\cmsorcid{0000-0002-3380-473X}, I.~Mikulec\cmsorcid{0000-0003-0385-2746}, J.~Schieck\cmsAuthorMark{2}\cmsorcid{0000-0002-1058-8093}, R.~Sch\"{o}fbeck\cmsAuthorMark{2}\cmsorcid{0000-0002-2332-8784}, D.~Schwarz\cmsorcid{0000-0002-3821-7331}, M.~Sonawane\cmsorcid{0000-0003-0510-7010}, W.~Waltenberger\cmsorcid{0000-0002-6215-7228}, C.-E.~Wulz\cmsAuthorMark{2}\cmsorcid{0000-0001-9226-5812}
% ...
\par}
\vskip\cmsinstskip
\dag:~Deceased\\
$^{1}$Also at Yerevan State University, Yerevan, Armenia\\
$^{2}$Also at TU Wien, Vienna, Austria\\
% ...
\end{sloppypar}
\end{document}